\documentclass[journal]{IEEEtran}
\usepackage{cite}
\usepackage{amsmath}
\usepackage{graphicx}
\usepackage{array}

\ifCLASSOPTIONcompsoc
 \usepackage[caption=false,font=normalsize,labelfont=sf,textfont=sf]{subfig}
\else
 \usepackage[caption=false,font=footnotesize]{subfig}
\fi
\usepackage[left]{lineno}
\usepackage{adjustbox}
\usepackage{xcolor}
\usepackage{hyperref}
\usepackage{epsfig}
\usepackage{multirow}

\DeclareMathOperator*{\argmax}{arg\,max}

\begin{document}

\title{A Workload Adaptive Haptic Shared Control Scheme for Semi-Autonomous Driving}
\author{Ruikun Luo*, Yifan Weng*, Yifan Wang, Paramsothy Jayakumar, Mark J. Brudnak, Victor Paul,\\ Vishnu R. Desaraju, Jeffrey L. Stein, Tulga Ersal$^{\dagger}$, and X. Jessie Yang$^{\dagger}$% <-this % stops a space
\thanks{We acknowledge the technical and financial support of the Automotive Research Center (ARC) in accordance with Cooperative Agreement W56HZV-19-2-0001 U.S. Army Ground Vehicle Systems Center (GVSC) Warren, MI.}
\thanks{R. Luo is with Robotics Institute, University of Michigan, Ann Arbor, MI.}
\thanks{Y. Weng, J.L. Stein, and T. Ersal are with the Department of Mechanical Engineering, University of Michigan, Ann Arbor, MI.}
\thanks{Y. Wang is with the Electrical Engineering and Computer Science, University of Michigan, Ann Arbor, MI.}
\thanks{P. Jayakumar, M.J. Brudnak, and V. Paul are with the U.S. Army Ground Vehicles System Center, Warren, MI.}
\thanks{V.R. Desaraju is with Toyota Research Institute, Ann Arbor, MI.}
\thanks{X.J. Yang is with the Department of Industrial and Operations Engineering, University of Michigan, Ann Arbor, MI.}
\thanks{* These authors have contributed equally. $\dagger$ These authors have contributed equally. }}
\maketitle
\begin{abstract}
    Haptic shared control is used to manage the control authority allocation between a human and an autonomous agent in semi-autonomous driving. Existing haptic shared control schemes, however, do not take full consideration of the human agent. To fill this research gap, this study presents a haptic shared control scheme that adapts to a human operator's workload, eyes on road and input torque in real-time. We conducted human-in-the-loop experiments with 24 participants. 
    In the experiment, a human operator and an autonomy module for navigation shared the control of a simulated notional High Mobility Multipurpose Wheeled Vehicle (HMMWV) at a fixed speed. At the same time, the human operator performed a target detection task for surveillance. The autonomy could be either adaptive or non-adaptive to the above-mentioned human factors. Results indicate that the adaptive haptic control scheme resulted in significantly lower workload, higher trust in autonomy, better driving task performance and smaller control effort. 
\end{abstract}

% \fi
% \IEEEpeerreviewmaketitle

\section{INTRODUCTION}
Autonomous driving technology is currently limited in its scope and reliability, giving rise to the semi-autonomous mode of driving.
In this mode, the driving task is shared between the human and the autonomy. 
Thus, properly allocating the control authority between these two agents becomes critical for safety and performance.
Managing this allocation is a challenging problem referred to as the shared control problem.

As the literature review in Sec. \ref{sec:literatureReview-SharedControl} shows, researchers have recognized this challenge and proposed many strategies for managing control authority in shared control. 
However, one very important consideration has been overlooked in the strategies developed to date; namely, the human operator's workload. 
The human operator's workload may change over time as the driving conditions change, or as a result of the human getting involved in secondary tasks. 
These variations in workload have direct implications on the management of control authority, as the human operator may or may not be ready to seize or relinquish control depending on the current workload.
However, the relationship between human workload and control authority management has not yet been explored.
This work aims to fill this gap by developing a shared control scheme that adapts to the human operator's workload in real-time.

\subsection{Background on shared control}
\label{sec:literatureReview-SharedControl}
Based on the hierarchy of authority, prior literature on shared control can be broadly classified into two categories: supervisory and co-operative.  

In supervisory shared control, one agent supervises the behavior of the other agent and determines the final control input to the vehicle. One example is a vehicle with Level~2 automation \cite{SAEAutonomy}, where the human acts as the supervisor. Human operator monitors the status of the vehicle and decides when to engage/disengage the automated driving function. On the other hand, in some schemes, the autonomy serves as the supervisor to monitor human operator's control commands and modify them as needed \cite{Erlien2016env, Schwarting2017, Storms2017}. In supervisory shared control, the control authority transfers entirely from one agent to the other in a discrete manner. Therefore, only one of the agents has the final control of the vehicle.

In co-operative shared control, both agents can affect the final control input. One type of co-operative shared control directly blends the steering angle inputs from both the human and autonomy through a designed arbitrator \cite{Anderson2011}. This scheme has the property that the loop between the human and the autonomy is closed after the steering wheel; i.e., the human will be able to feel the impact of autonomy input only after the resultant steering command takes effect and through the response of the vehicle. The other type of co-operative shared control is haptic shared control, in which the human and autonomy can negotiate the steering angle through the torques they apply to the steering wheel \cite{brent2005, sebastiaan2015, Nguyen2018attention}. In this scheme, the human operator can directly feel the torque from the autonomy and can choose to yield to or fight with it by exerting extra torque on the steering wheel. Researchers developed and tested a haptic shared control framework, and showed that haptic control improved the driving performance while reducing visual demand or shortening the reaction time of the secondary task \cite{brent2005}. Others used the haptic control framework with the bandwidth guidance version and the continuous guidance version, and showed that both helped reduce driver errors \cite{sebastiaan2015}. 

% \textcolor{red}{Jessie: @Yifan, here we need to point out the shortcomings of the traditional haptic shared control scheme. And state that we are proposing an adaptive haptic shared control. The proposed control scheme will adapt to workload, attention, and torque. }

The impedance of autonomy in a haptic shared control scheme can be considered as a natural tuning parameter through which adaptability can be introduced. Indeed, even though earlier haptic shared control schemes used a fixed impedance \cite{brent2005, abbink2008}, later works started investigating adaptive impedance schemes. Some schemes adopt vehicle-performance-based switching rules as adaptation mechanisms, such as turning shared control on when the lateral error of the vehicle exceeds a designed threshold \cite{sebastiaan2015}.
Others consider human-performance-based metrics to adapt impedance continuously, such as human's input torque and attention as the guideline for designing control authority allocation \cite{Nguyen2018attention}. 
However, workload, an important human factor, has not yet been considered for adaptation purposes.

Therefore, in the present work, we proposed an adaptive haptic shared control scheme, which considers the human operator's workload, eyes on road and torque input in the control authority allocation.

\subsection{Background on workload estimation}
Workload can be measured offline or online. Offline retrospective measures are commonly used after a human operator completes a task via questionnaires such as the NASA TLX (Task Load Index) \cite{hart1988development}. Their offline nature prohibits their utilization for real-time adaptation.
Online real-time measures of workload are assessed while the human operator is performing the task and thus could be used for designing adaptive systems. Online real-time measures of workload are usually based on task performance or human operator's physiological signals. The underlying rationale for performance-based measures is that under high workload, human operator's task performance would be harmed. Performance-based measures, however, are not applicable if the task performance is ambiguous or is not available immediately. Physiological measures rely on changes in the human physiological signals. Various types of physiological signals have been used to estimate workload, including heart rate, electroencephalogram (EEG), eye-related measures, Galvanic Skin Response (GSR) and near infrared spectroscopy (NIRS). Please refer to \cite{Heard:2018ip} for a review. 

Among all the physiological measures, some could be intrusive  (e.g., EEG \cite{liu2017multisubject}) or could be easily affected by body movements (e.g., heart rate \cite{chen2015wavelet}). 
Eye-tracking emerges as a less intrusive and robust technique and research efforts have been spent on using eye-related measurements to assess operators' workload, including pupil diameter \cite{recarte2003mental, lu2019workload}, gaze distribution \cite{reimer2009impact}, gaze trajectory \cite{wang2014sensitivity, Fridman2018}, and blink rate \cite{coral2016analyzing}.

To assess workload online using physiological measurements, previous studies largely adopted statistical methods to show the relationships between certain physiological signals and workload. Recently, researchers started to use machine learning techniques to classify mental workload into different levels. For example, some researchers used a decision tree to classify the drivers' workload into two levels using 30~s driving data and the pupil diameter data \cite{zhang2004driver}. Others proposed a deep neural network to analyze a 6~s video of the eye and classified operators' workload into 3 categories in real-time \cite{Fridman2018}. However, such online measures of workload have not yet been incorporated into the shared control schemes.

\section{PRESENT STUDY}
In the present study, we developed a dual-task shared control platform. Using the experimental platform, the human operator and the autonomy shared the control of a simulated notional High Mobility Multipurpose Wheeled Vehicle (HMMWV) at a fixed speed. At the same time, the human operator performed a target detection task for surveillance. We used the Hidden Markov Model (HMM) to estimate the human's workload by analyzing 4~s gaze trajectory data. We then designed and tested the adaptive shared control scheme by regulating the assistance level of the autonomy based on the estimated workload, and the operator's eyes on road and input torque.

\subsection{Workload estimation with HMM}
\label{sec:workloadEstimation}
% \subsubsection{}
% HMM has been used to model gaze trajectory to estimate workload \cite{Fridman2018}. 

\begin{figure}[t]
\centering
\includegraphics[width = 0.6\linewidth]{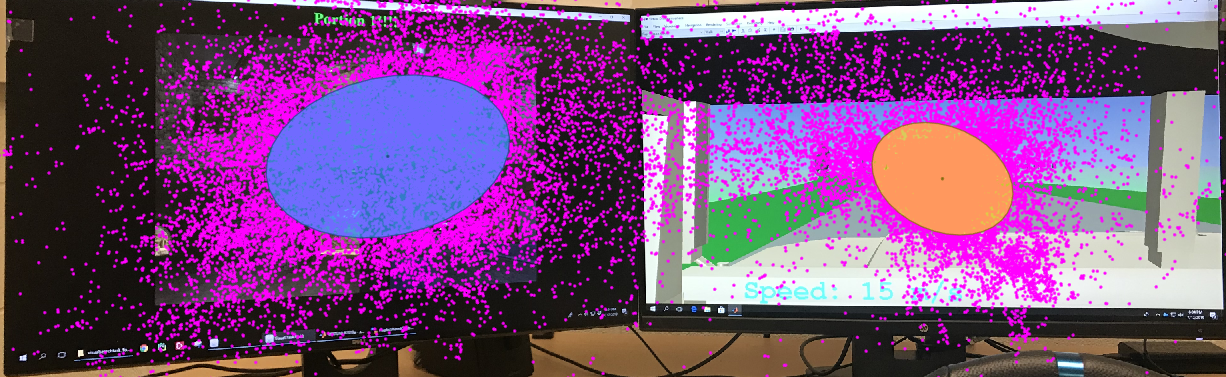}
    \vspace{-3mm}
\caption{Example of using the Hidden Markov Model to model gaze trajectory to estimate workload. Magenta dots: gaze points. Ellipsoids: Multivariate normal distributions.}
\label{fig:hmm_example}
\end{figure}

HMM contains a set of hidden states $S = \{s_1, s_2, ..., s_N\}$, where $N$ is the number of hidden states, time sequence observations $o_t$, observation model $p(o_t|s_j)$, and state transition probabilities $p(s_i|s_j)$. In the present study, the time sequence observations are the gaze points, i.e., locations of where the human is looking at relative to the external world coordinate. Let $\mathbf{O} = \{\mathbf{o_1}, \mathbf{o_2}, ..., \mathbf{o_T}\}$ represent a gaze trajectory captured from the eye tracker, where $o_t$ represents the gaze point at time $t$. Therefore, the hidden states are centers of the gaze points, $\mu_1, ...\mu_N$ and the observation model is the multivariate normal distributions over centers, $p(\mathbf{o_t}|s_j) \sim \mathcal{N}(\mu_j,\Sigma_j)$ as shown in Fig.~\ref{fig:hmm_example}.

We trained two HMMs, one for the high workload and one for the moderate workload. The parameters of the HMMs were learned by the Expectation Maximization algorithm using the open source implementations from \cite{rozo2016learning, Calinon15}. The number of hidden states was determined by the Bayesian Information Criterion (BIC) \cite{calinon2005recognition, schwarz1978estimating}. 

Given a gaze trajectory $X$, we computed the likelihood of $p(X|H_i)$ via the forward algorithm, where $H_i$ represents different learned HMMs for the high workload and moderate workload. To estimate the workload of $X$, we found the HMM with the maximum likelihood, i.e., $\argmax\limits_i p(X|H_i)$.

% \subsubsection{Real time data streaming}
Our proposed adaptive shared control scheme is based on the human operator's real-time workload, eyes on road and input torque. We used the gaze point data from a 4~s time window captured by the Tobii eye tracker (30~Hz sampling rate) to estimate participants' workload and eyes on road. Thus, $T = 120$. 
Let $w_t$ represent a human operator's workload at time $t$, $w_t = c_1 \argmax\limits_i p(\mathbf{O_t}|H_i) + c_2$, where $c_1, c_2$ are scaling and offset factors such that $w_t = 50$ represents moderate workload, and $w_t = 100$ represents high workload. A human operator's eye on road is defined as the percentage of time that s/he is looking at the driving task. Let $e_t$ denote the human operator's eyes on road. $e_t$ is calculated as the average number of times that a participant's gaze points fall on the driving screen within the time window $T$. 

%  Let $\mathbf{O_t} = \{\mathbf{o_{t-T+1}}, \mathbf{o_{t-T+2}}, ..., \mathbf{o_t}\}$ represent the 4~s time window gaze trajectory at time $t$, where $\mathbf{o_t} = (x_t, y_t)$. Define the function $f(\mathbf{o_t})$ such that $f(\mathbf{o_t}) = 1$ if $(x_t, y_t)$ is within the driving screen; $f(\mathbf{o_t}) = 0$ otherwise. Thus, $e_t = \frac{\sum_{i=t-T+1}^{t}f(o_i)}{T}$.

Due to the large mass and high center of gravity of the simulated military vehicle (see Sec.~\ref{sec:Experiment_1_method}), a rapid change of control commands resulting from a rapid change of $w_t$ and $e_t$ could trigger a rollover. Therefore, we applied a moving average filter with a 1~s time window and downsampled $w_t$ and $e_t$ to 10~Hz.

\subsection{Adaptive shared control scheme}
\subsubsection{Autonomy design} 
In the present work, we designed a fixed speed scenario when the vehicle travels at 15 mph. Hence, the autonomy only needed to provide the steering angle commands for reference. We used the Nonlinear Model Predictive Control (NMPC) method to generate the steering wheel commands, which can track the given centerline for the vehicle. Refs. \cite{liu2017mpc} and \cite{Febbo2017} describe the formulation of the NMPC in detail. We used the same bicycle model representation of the vehicle within the NMPC framework with the same states and control constraints as \cite{liu2017mpc,Febbo2017}. We tailored the cost function to fit to our problem, because unlike the scenarios for which the original NMPC formulation was developed, we limited the vehicle traveling speed to 15 mph and there were no obstacles on the path. Specifically, the cost function was defined as 
% \begin{equation}
%     \begin{split}
%         J = & w_1 \int_{t_0}^{t_P} \Big(y_\textrm{ref}(x(t)) - y(t)\Big)^2 dt + w_2 \int_{t_0}^{t_P} \gamma^2 dt \crcr 
%         &+ w_3 \int_{t_0}^{t_P} \tanh{\Big[-\frac{F_{z,rl}-a }{b}\Big]} + \tanh{\Big[-\frac{F_{z,rr}-a }{b}\Big]}  dt
%     \end{split}
% \end{equation}

\begin{equation}
    \begin{split}
        J = & w_1 \int_{t_0}^{t_P} \Big(y_\textrm{ref}(x(t)) - y(t)\Big)^2 dt + w_2 \int_{t_0}^{t_P} \gamma^2 dt \crcr 
        &+ w_3 \int_{t_0}^{t_P} \tanh{\Big[\frac{a-F_{z,rl}}{b}\Big]} + \tanh{\Big[\frac{a-F_{z,rr}}{b}\Big]}  dt
    \end{split}
\end{equation}

The cost function comprises three terms. The first term 
% $w_1 \int_{t_0}^{t_P} (y_\textrm{ref}(x(t)) - y(t))^2 dt$ 
is designed to penalize the distance from the position of the vehicle $y(t)$ to the given position on the centerline $y_\textrm{ref}(x(t))$. 
The second term 
% $w_2 \int_{t_0}^{t_P} \gamma^2 dt$ 
is designed to regularize the steering rate $\gamma$, which ensures the steering angle command changes smoothly. 
The third term 
% $w_3 \int_{t_0}^{t_P} \tanh{\Big[-\frac{F_{z,rl}-a }{b}\Big]} + \tanh{\Big[-\frac{F_{z,rr}-a }{b}\Big]}  dt$ 
is a soft constraint that increases the cost when one of the tire vertical loads $F_{z,rl}$, $F_{z,rr}$ is close to the lowest allowable threshold. This soft constraint is used to prevent the vehicle from operating at its dynamic limit unnecessarily \cite{liu2017mpc,liu2018mpc}. The weights $w_1$, $w_2$ and $w_3$ are set to achieve a trade-off between the three terms in the cost function. In this formulation, $t_0$ is the time when the prediction horizon starts, while $t_p$ marks the end time of the prediction horizon. $t_p = t_0 + T_p$, where $T_p$ is the fixed prediction horizon and it equaled 6.5~s in this work. We used the open-source  nonlinear optimal control package NLOptControl \cite{nlopt2017}, which uses the Legendre-Gauss-Radau collocation method to transfer the continuous optimal control problem into a nonlinear program. We then solved the nonlinear program by using the solver package IPOPT \cite{Wachter2006}. This optimization process  generates a series of steering angle commands through the whole control horizon $T_p$, and we use only the first 3~s worth of commands. While executing the previous control command series, the system formulates and solves a new optimal control problem with a receded horizon, and the resulting new command series are applied as soon as they are available.

\subsubsection{Non-adaptive haptic shared control}
\label{sec:nonAdaptiveSharedControl}
\begin{figure}
    \centering
    \includegraphics[width = 0.7\linewidth]{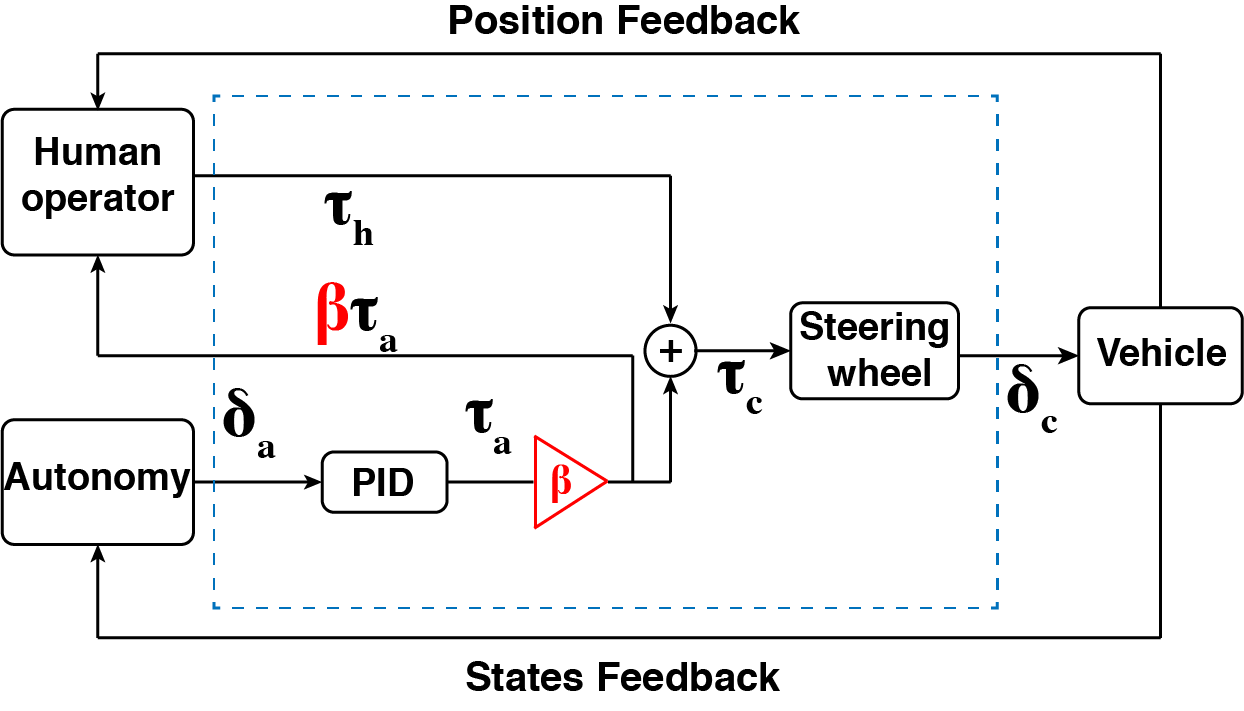}
        \vspace{-3mm}
    \caption{Block diagram for haptic shared control. $\delta_a$ represents the steering angle command from autonomy, $\tau_h$ and $\tau_a$ represent the torque from human and autonomy, respectively. $\tau_c$ and $\delta_c$ are the actual control torque and actual control steering angle. $\beta$ is the assistance level, which is always 1 in the baseline non-adaptive scheme, whereas it varies in the proposed adaptive scheme.}
    \label{fig:two control diagrams}
\end{figure}

Haptic shared control combines the torques applied by the autonomy and human operator. It creates a smooth control authority transfer between the human operator and autonomy. The implementation is visualized in Fig.~\ref{fig:two control diagrams}, where $\beta=1$ for the baseline non-adaptive case.  

The torque from the autonomy comes from a proportional-integral-derivative (PID) controller, which acts on the difference between the steering commands resulting from the NMPC framework as the reference trajectory and the current steering angle measurement.
When there is no input from the human operator, the autonomy follows the reference centerline it perceives. The perceived reference centerline may be different from the actual centerline. When there is an input from the human operator that deviates the vehicle from the centerline autonomy perceives, the autonomy applies extra torque to bring the vehicle back to the perceived centerline. The human operator can hence feel the intention of the autonomy and decide whether s/he would agree with it and let autonomy have more control authority (yield), or claim more control authority (fight). The resultant torque applied on the steering wheel, which is the summation of the torques from the human operator and autonomy, determines the final steering angle applied to the vehicle.

\subsubsection{Adaptive haptic shared control}
We designed our adaptive shared control scheme based on three different features: workload, torque from the human operator, and eyes on road. The resultant torque $\tau_c$ in the adaptive scheme is $\tau_c = \tau_h + \beta(w_t, e_t, \hat{\tau}_h) \tau_a$, where the term $\beta$ is referred to as assistance level and it determines the strength of assistance torque from autonomy. $\hat{\tau}_h$ is the normalized human torque that is calculated by dividing the input torque from the human operator by the maximum torque a human operator can apply.
The implementation of the adaptive scheme is shown in Fig.~\ref{fig:two control diagrams}. This scheme is in contrast to the direct blending of the input torques from both the human operator and autonomy as in the non-adaptive haptic shared control scheme. Specifically, $\beta$ is always 1 in the baseline non-adaptive haptic shared control scheme, whereas it varies in the proposed adaptive scheme. 

In our heuristic design for the assistance level, $\beta$ was separated into two parts: base assistance level $\Bar{\beta}$ and assistance level increment $\Delta{\beta}$; i.e., $\beta = \Bar{\beta}(w_t, \hat{\tau}_h) + \Delta{\beta}(w_t, e_t)$. The base assistance level $\Bar{\beta}$ considers the impact from workload and input torque from the human operator, while the assistance level increment $\Delta{\beta}$ considers the combined effect of eyes on road and workload.  

\begin{figure}[t]
\centering
\subfloat[Relationship between base assistance level $\bar{\beta}$ and workload $w_t$\label{fig:assistance_workload}]{\includegraphics[width = 0.45\linewidth]{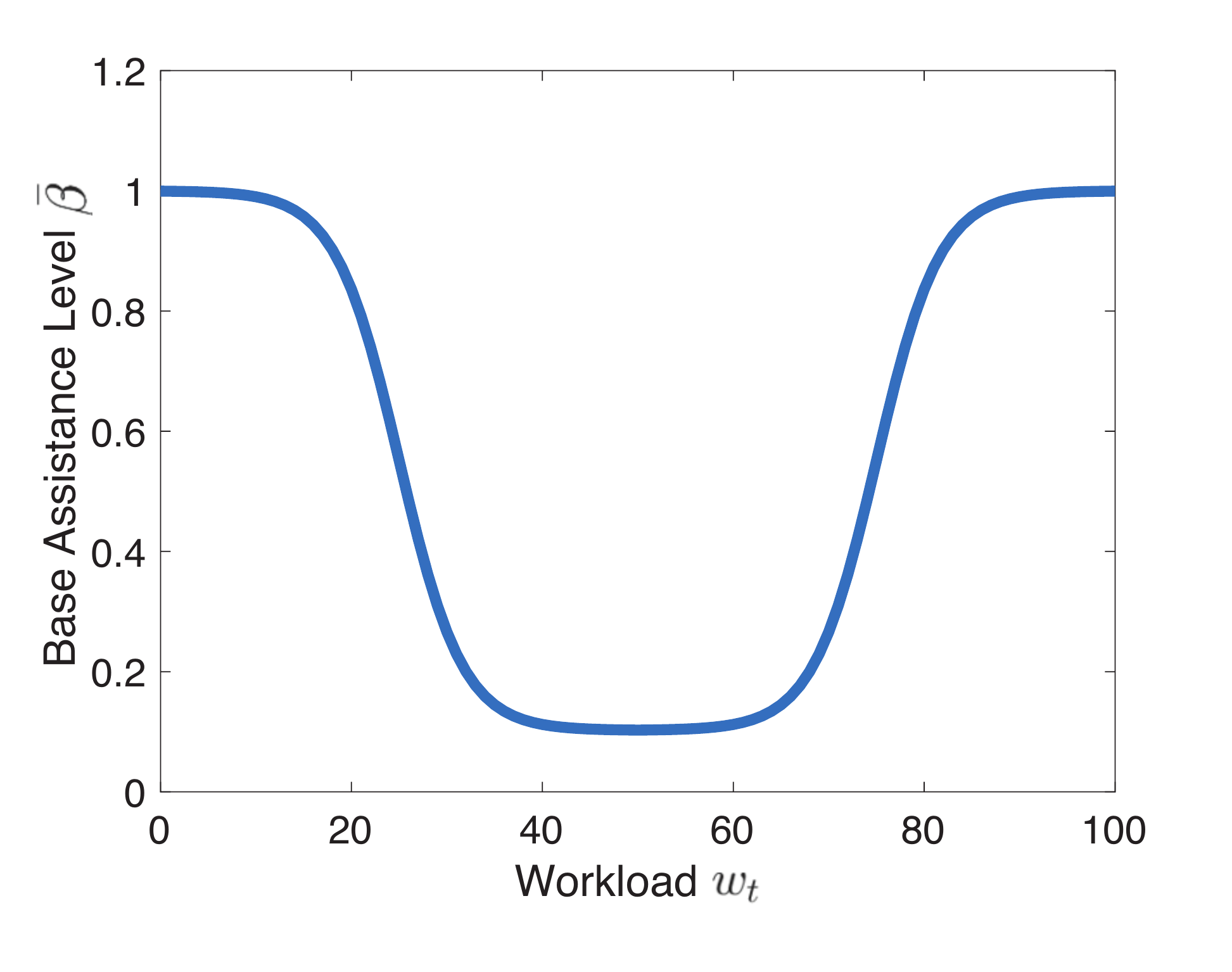}}
\quad
\subfloat[Relationship between base assistance level $\bar{\beta}$ and normalized human torque $\hat{\tau}_h$\label{fig:assistance_torque}]{\includegraphics[width = 0.45\linewidth]{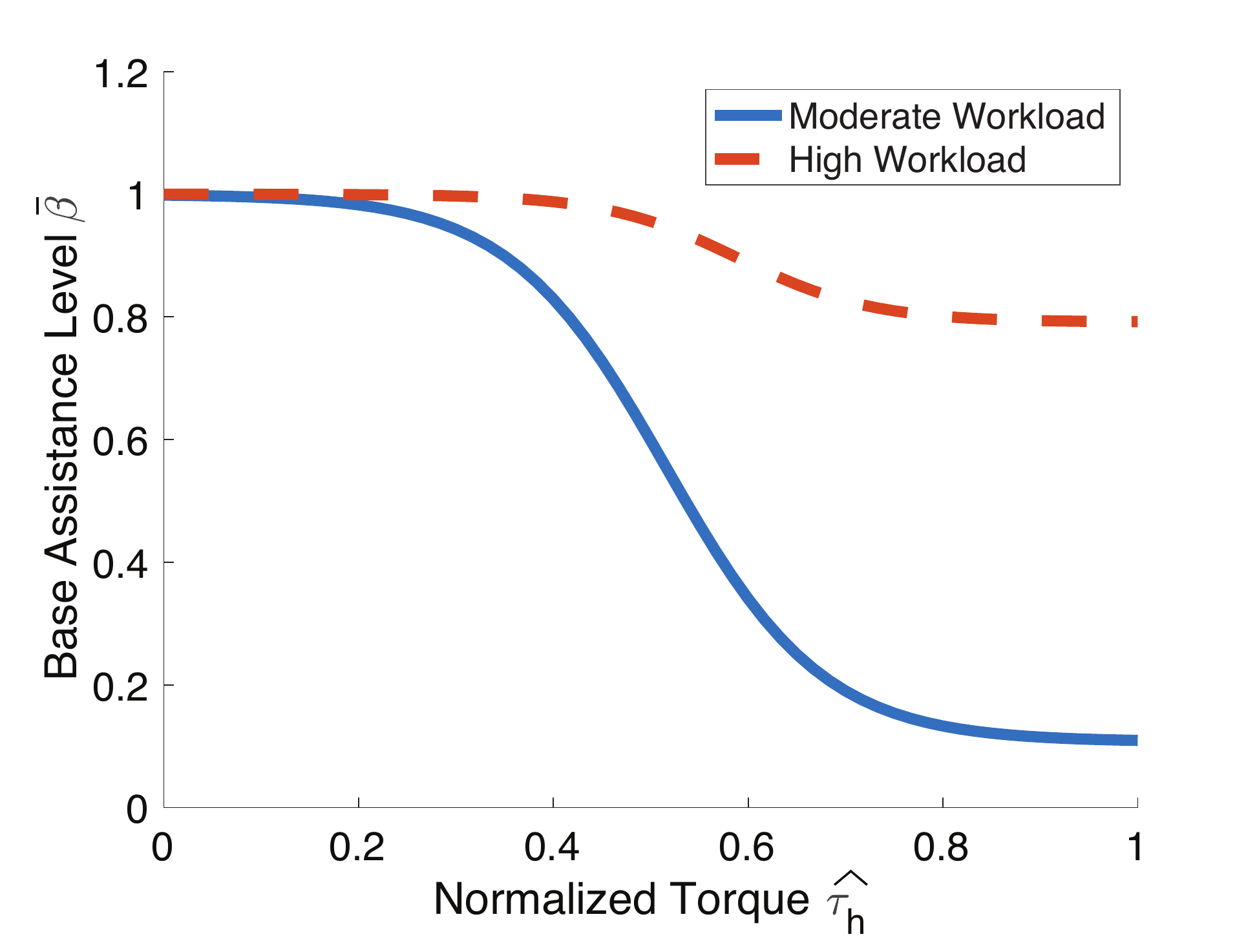}}
    \vspace{-2mm}
\caption{Illustration of base assistance level design principles}
\label{fig:base_principles}
\end{figure}

The base assistance level $\Bar{\beta}$ was designed according to the principles illustrated in Fig.~\ref{fig:base_principles} and explained next.
On the one hand, when the torque from the human operator is held constant, the relationship between the assistance level and workload is shown in Fig.~\ref{fig:assistance_workload}. The designed curve for assistance level matches the study of \cite{Flemisch2010TowardsHA}, which shows the assistance from the autonomy should be high when the workload is either very high (overloaded) or very low (underloaded). When the subject has a moderate workload, the assistance from the autonomy should be lower. We set the workload value $w_t$ as 0 when the subject is underloaded, $w_t$ as 50 when the subject experiences moderate workload and $w_t$ as 100 when the subject is overloaded. We heuristically set the assistance level as 0.1 for moderate workload ($w_t = 50$) and as 1 for very high workload ($w_t = 100$). We fit a sigmoid function to create the smooth transition from $w_t = 50$ to $w_t = 100$. We then mirror the function when workload $w_t$ ranges from 0 to 50 and obtain the curve for the whole workload spectrum.

On the other hand, when the workload of the human operator is held constant, the relationship between the assistance level $\beta$ and normalized human torque $\hat{\tau}_h$ is shown in Fig.~\ref{fig:assistance_torque}. There are two critical properties of the designed curve. When the human torque is small, the assistance level is kept at a high level ($\Bar{\beta}$ = 1) to filter out some unintended input torque from the human operator. The assistance level starts to drop after the normalized human torque passes a threshold, which increases as workload increases from moderate workload ($w_t = 50$) to high workload ($w_t = 100$). The threshold value from underloaded to moderate workload mirrored the threshold value when the workload ranges from moderate to high. The threshold is smaller when workload approaches the moderate level, since we assume human would make less mistakes at this workload level based on the results in the literature that show a moderate workload level to be optimal \cite{Flemisch2010TowardsHA}. We heuristically set the threshold value as 0.01 when the human operator experiences a moderate workload ($w_t = 50$), while this value is 0.3 when the human operator is fully overloaded and underloaded ($w_t = 100$ and $w_t = 0$). We fit a quadratic function that is symmetric about the moderate workload case ($w_t = 50$). When the input torque from the human operator becomes greater, the assistance level starts to drop to a lower level. The system is designed to let the human operator have more control authority when there is a strong intention for intervention from the human operator. The value also changes according to the workload. When the human is overloaded, the assistance level for a large torque input remains a relatively large value. We heuristically set the assistance level for maximum torque as 1 when the human operator is overloaded ($w_t = 100$) and 0.1 when the human operator experiences moderate workload ($w_t = 50$). We then use a modified sigmoid function, connecting the threshold point with the maximum torque point. 

\begin{figure}[t]
\centering
\includegraphics[width = 0.6\linewidth]{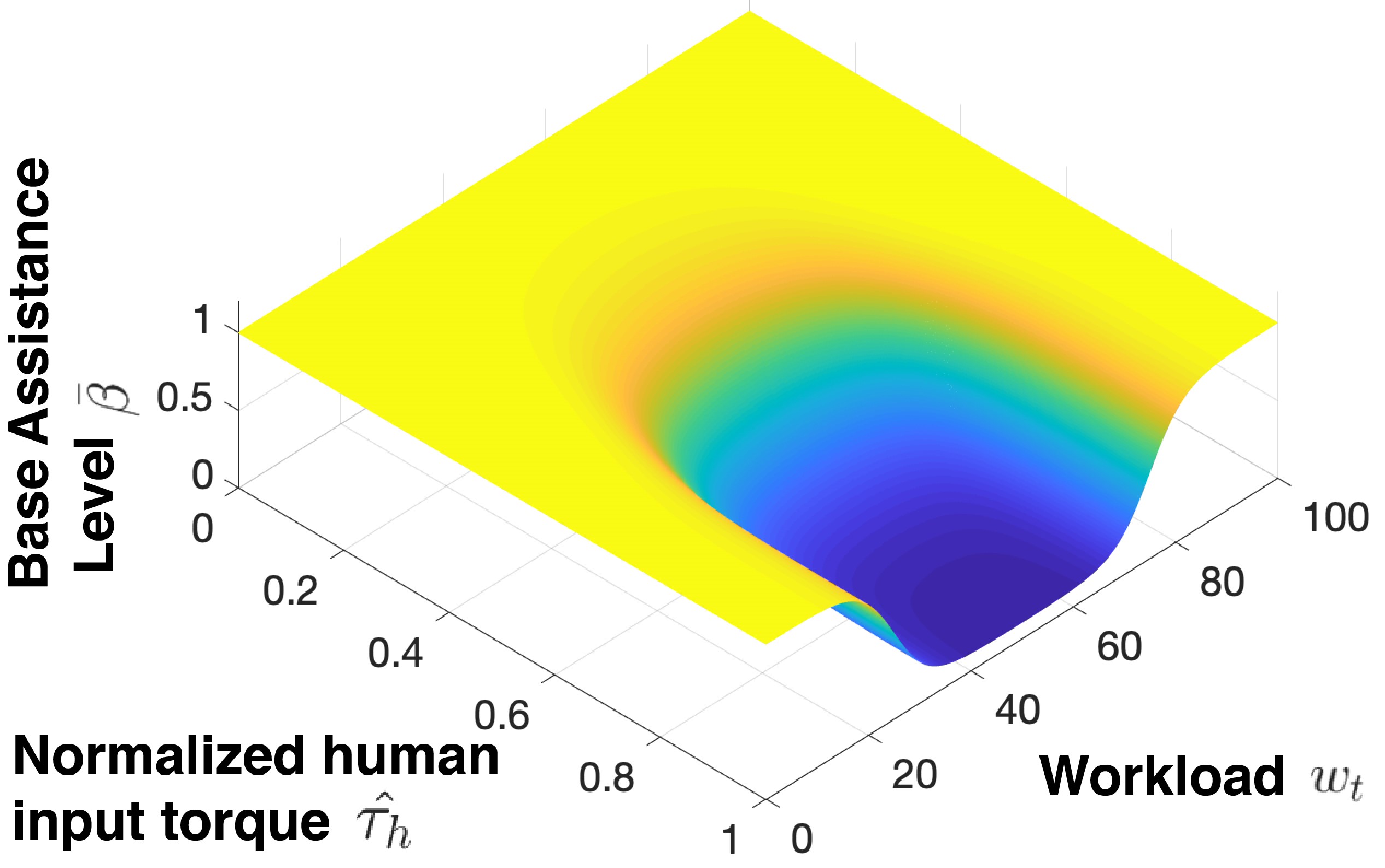}
    \vspace{-3mm}

\caption{Relationship between base assistance level $\Bar{\beta}$, workload $w_t$, and normalized human input torque $\hat{\tau}_h$}
\label{fig:assistance_3d}
\end{figure}

Combining those two principles, the formulation of the base assistance level $\Bar{\beta}$ is obtained as

\begin{equation}
    \begin{split}
        \Bar{\beta}(w_t, \hat{\tau}_h) = &1-\Big [1-(\frac{0.9e^{0.3(|w_t-50|-25)}}{e^{0.3(|w_t-50|-25)}+1}+0.1)\Big]\crcr
        &\Bigg[\frac{e^{\frac{72\hat{\tau}_h-36.6-15(\frac{w_t}{50}-1)^2}{5.9-2.5(\frac{w_t}{50}-1)^2}}}{e^{\frac{72\hat{\tau}_h-36.6-15(\frac{w_t}{50}-1)^2}{5.9-2.5(\frac{w_t}{50}-1)^2}}+1}\Bigg]
    \end{split}
\end{equation}
 The corresponding 3D plot showing the relationship between the baseline assistance level, the workload and the normalized human torque is shown in Fig.~\ref{fig:assistance_3d}.

\begin{figure}[t]
\centering
\subfloat[Relationship between assistance level increment $\Delta{\beta}$ and eyes on road for different workloads \label{fig:increment_eyes}]{\includegraphics[width = 0.45\linewidth]{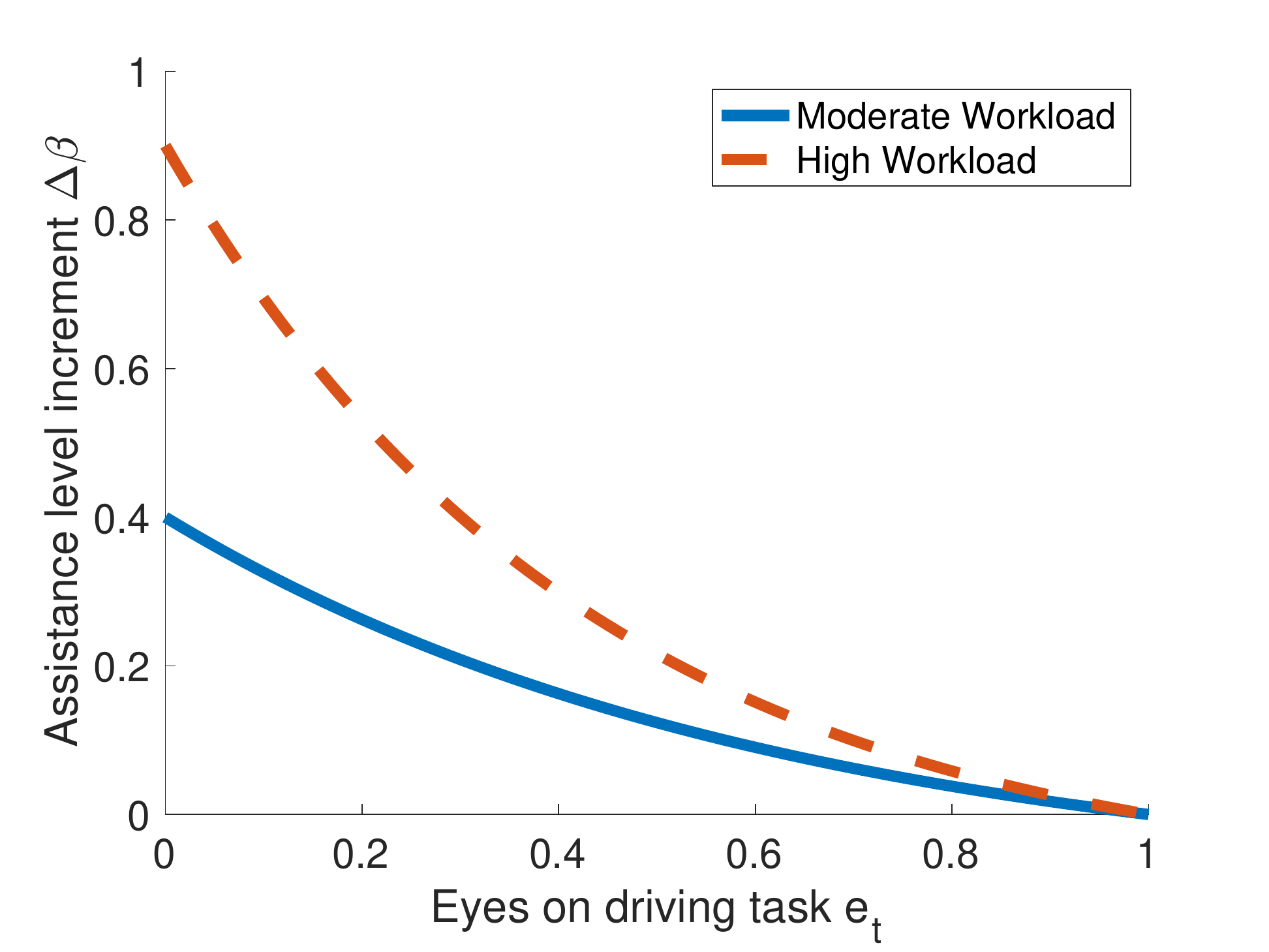}}
\quad
\subfloat[Relationship between assistance level increment $\Delta\beta$ and workload $w_t$ when $e_t$ = 0\label{fig:increment_workload}]{\includegraphics[width = 0.45\linewidth]{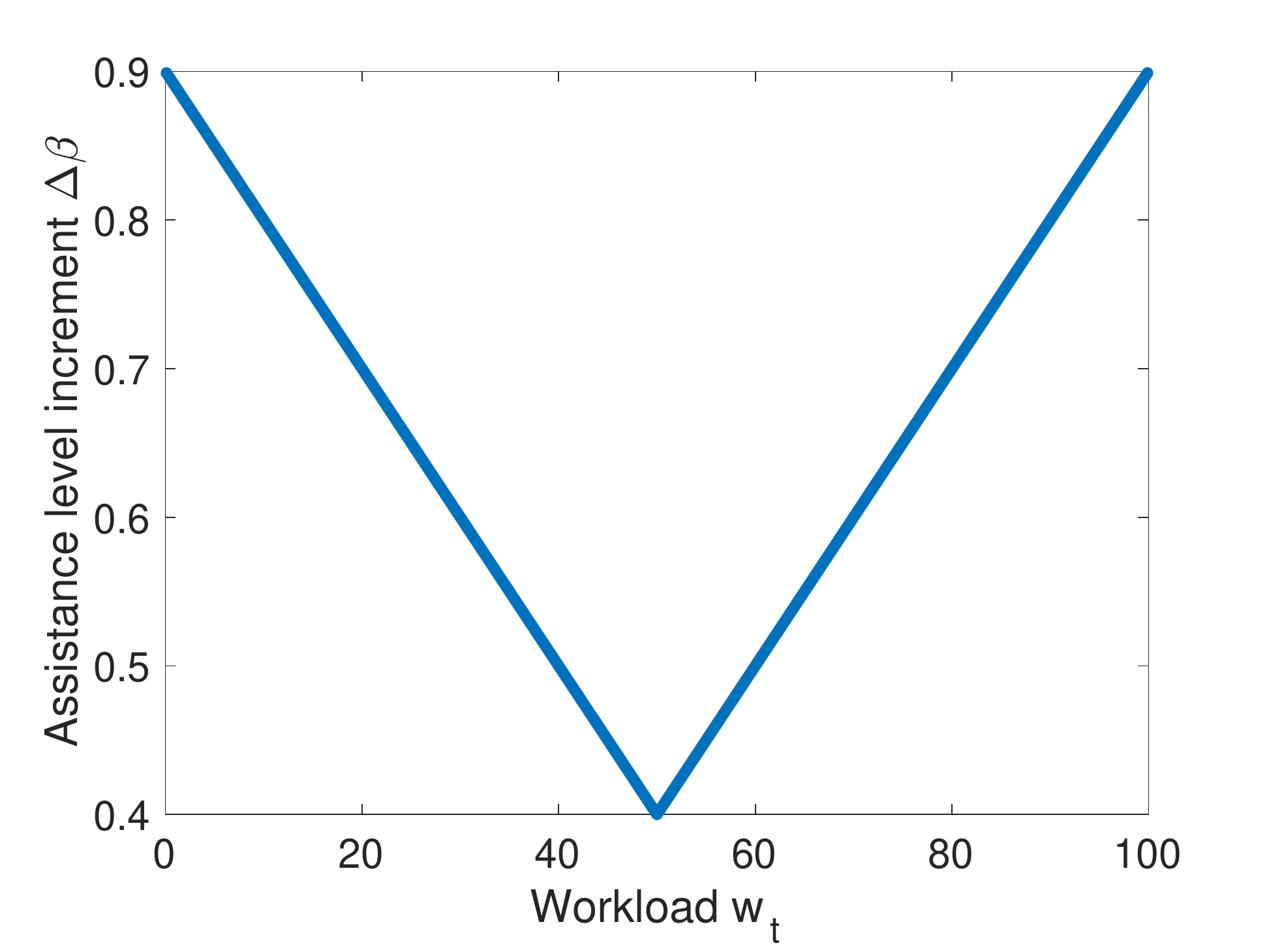}}
    \vspace{-2mm}
\caption{Illustration of assistance level increment design principles}
\label{fig:base_principles_increment}
\end{figure}

The assistance level increment $\Delta{\beta}$ was designed according to the principles illustrated in Fig.~\ref{fig:base_principles_increment} and explained next.
On the one hand, keeping the workload constant, when the subject focuses on the driving task, i.e., $e_t$ is very close to 1, $\Delta{\beta}$ is very close to 0, which indicates no additional assistance level is provided based on the the eyes on road metric. When the subject directs their attention to the secondary tasks, i.e., $e_t$ is very close to 0, $\Delta{\beta}$ increases to a high level, which is illustrated in Fig.~\ref{fig:increment_eyes}. An exponential function is used to connect these two points. On the other hand, keeping the eyes on road constant, when workload is high, the increment $\Delta{\beta}$ is large, while when the workload is moderate, the increment $\Delta{\beta}$ is small, which is shown in Fig.~\ref{fig:increment_workload}. We heuristically set the value of $\Delta{\beta}$ as 0.4 when the subject experiences moderate workload ($w_t = 50$), and as 0.9 when the subject is overloaded or underloaded ($w_t = 100$ or $w_t = 0$). This value is calculated through linear interpolation when the workload is between these critical values.

\begin{figure}[t]
\centering
\includegraphics[width = 0.6\linewidth]{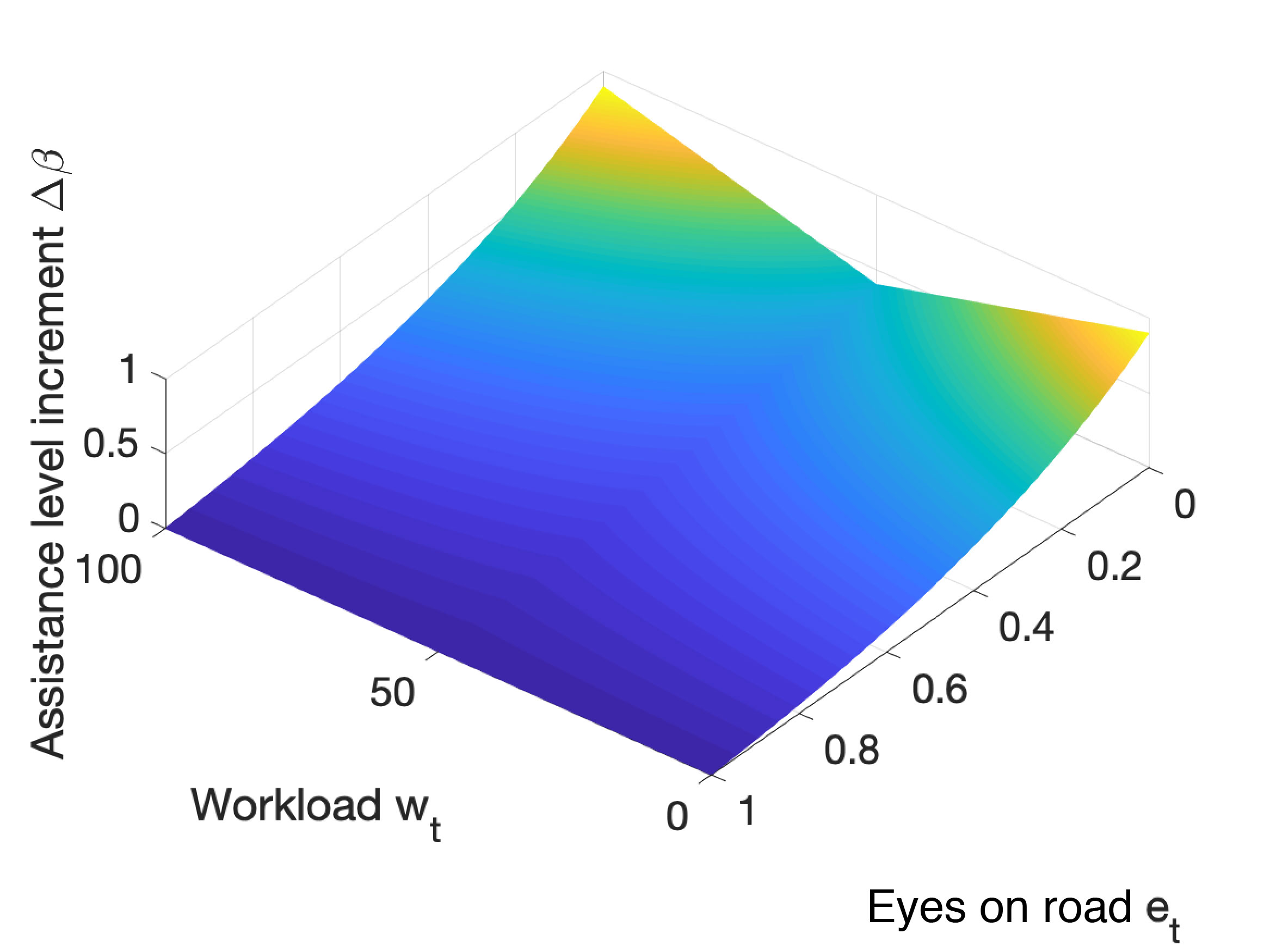}
    \vspace{-4mm}
\caption{Relationship between assistance level increment $\Delta\beta$,  workload $w_t$, and eyes on road}
\label{fig:assistance_increment}
\end{figure}

Combining these considerations, the formulation of assistance level increment $\Delta{\beta}$ is obtained as
\begin{equation}
    \Delta{\beta}(w_t, e_t) = 0.1(0.1|w_t-50|+5) ^{1-e_t} - 0.1
\end{equation}
The corresponding 3D plot showing the relationship between the assistance level increment, the workload and the eyes on road is shown in Fig.~\ref{fig:assistance_increment}. %The assistance level $\beta$ is a direct sum of the two terms $\bar{\beta}$ and $\Delta\beta$. 

%The assistance level $\beta$ is a direct sum of the two terms $\bar{\beta}$ and $\Delta\beta$. 

%We made two hypotheses when choosing the heuristic values used in the adaptive scheme. First, recall that when the subject experiences moderate workload and focuses on the driving task, we set the assistance level $\beta$ lower than 1. We hence expect a drop in human input torque when compared with the non-adaptive scheme. Second, because a previous study mentioned that when continuous guidance is provided, the mean lane keeping error is reduced when the impedance of the haptic system is large \cite{sebastiaan2015}, we increase assistance level $\beta$ to greater than 1 when the subject has high workload and focuses on the surveillance task. Therefore, we expect a decrease in lane keeping error when compared with the non-adaptive scheme. \Tulga{I do not fully understand the purpose of this paragraph. Is it to provide some explanation on how the heuristic values in the scheme were chosen? Or is it to provide the hypotheses we wanted to test in the experiments? If it is the former, then it is not doing a great job with explaining all the heuristic choices. If it is the latter, then I am not sure if this is the best place to list the hypotheses. In my mind, where a paragraph like this fits best is in the results section to explain some of the results we obtained instead of posing the statements as hypotheses.} \Yifan{We just want to provide the hypotheses we wanted to test in the experiments}

\section{EXPERIMENT 1}
\subsection{Introduction}
In Experiment 1, we aimed to estimate a human operator's workload in real-time by analyzing his/her gaze trajectories. We conducted a human-in-the-loop experiment with 12 participants using a dual-task shared control platform.  The participant and the autonomy shared the control of a simulated notional HMMWV. At the same time, the participant performed a surveillance task. The participant wore a pair of Tobii eyeglasses 2 during the experiment and his/her gazes were captured at 30~Hz. Based on the gaze trajectories, we estimated the participant's workload using HMM.

% This experiment is used to collect human's physiological data to evaluate the performance of workload estimation and learn parameters for Hidden Markov Models used in the Experiment 2 to evaluate the effect of adaptive shared control scheme on human's task performance.

% Prior to this Experiment 1, we conducted two pilot study to select tracks in the Experiment 1 and 2 and evaluate task difficulty for different surveillance task fixed time (see details in Appendix).

\subsection{Method}
\label{sec:Experiment_1_method}
% This study complied with the American Psychological Association code of ethics and was approved by the Institutional Review Board at the University of Michigan.

\subsubsection{Participants}
A total of 13 university students participated in the experiment. Data from one participant were discarded due to equipment malfunction. The remaining 12 participants were on average 26.7 years old (\textit{SD} = 3.0 years) and had an average of 8.3 years of driving experience (\textit{SD} = 4.4 years). All participants had normal or corrected-to-normal vision.
% and their eyes are able to be calibrated by the eye tracker.

\subsubsection{Apparatus and stimuli}
\begin{figure}[t]
\centering
\subfloat[Driving task (right screen) and surveillance task (left screen)]{\includegraphics[width = 0.45\linewidth]{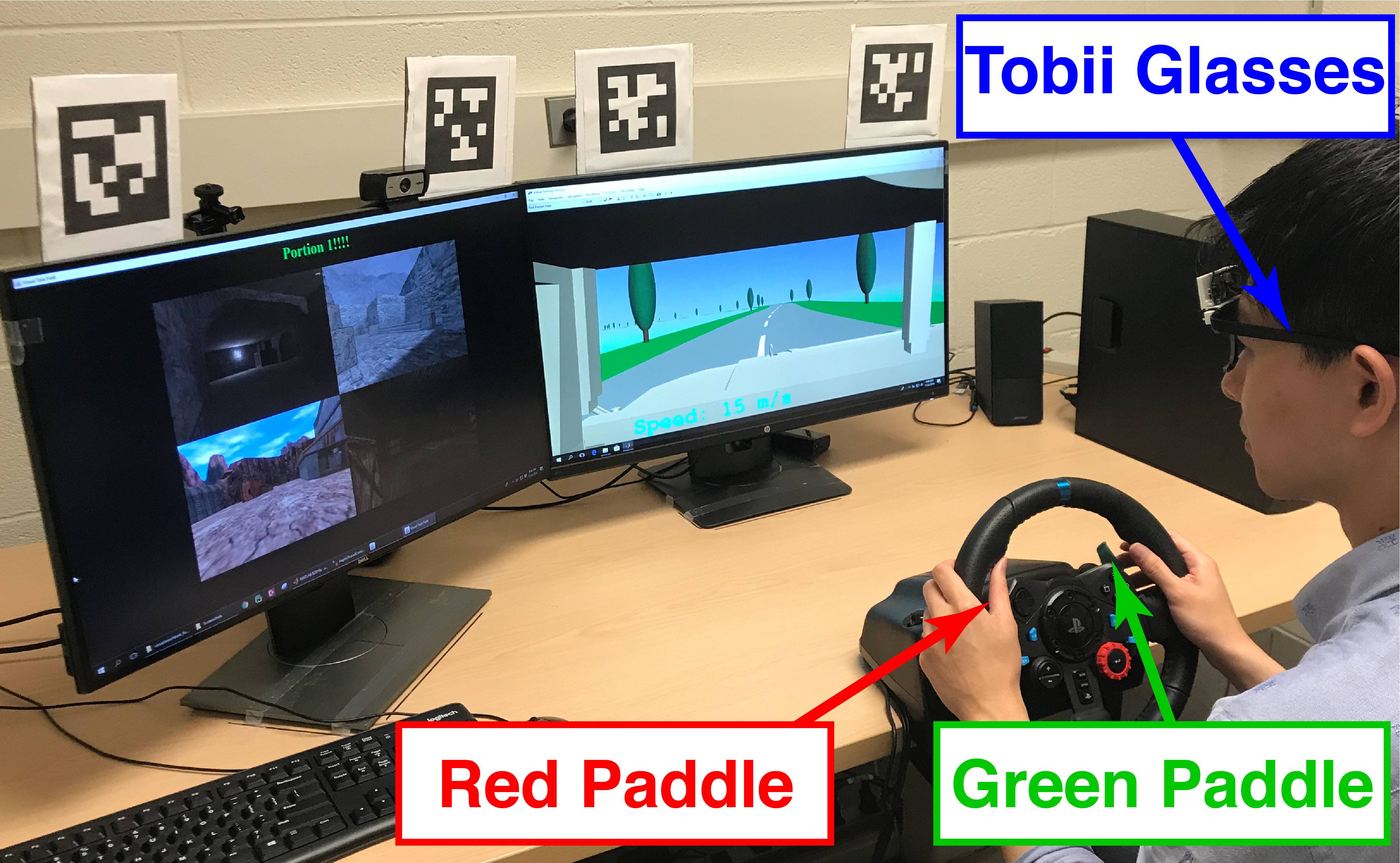}
    }
\quad
\subfloat[Red and green paddles used to perform the surveillance task \label{fig:testbed_pedal}]{\includegraphics[width = 0.45\linewidth]{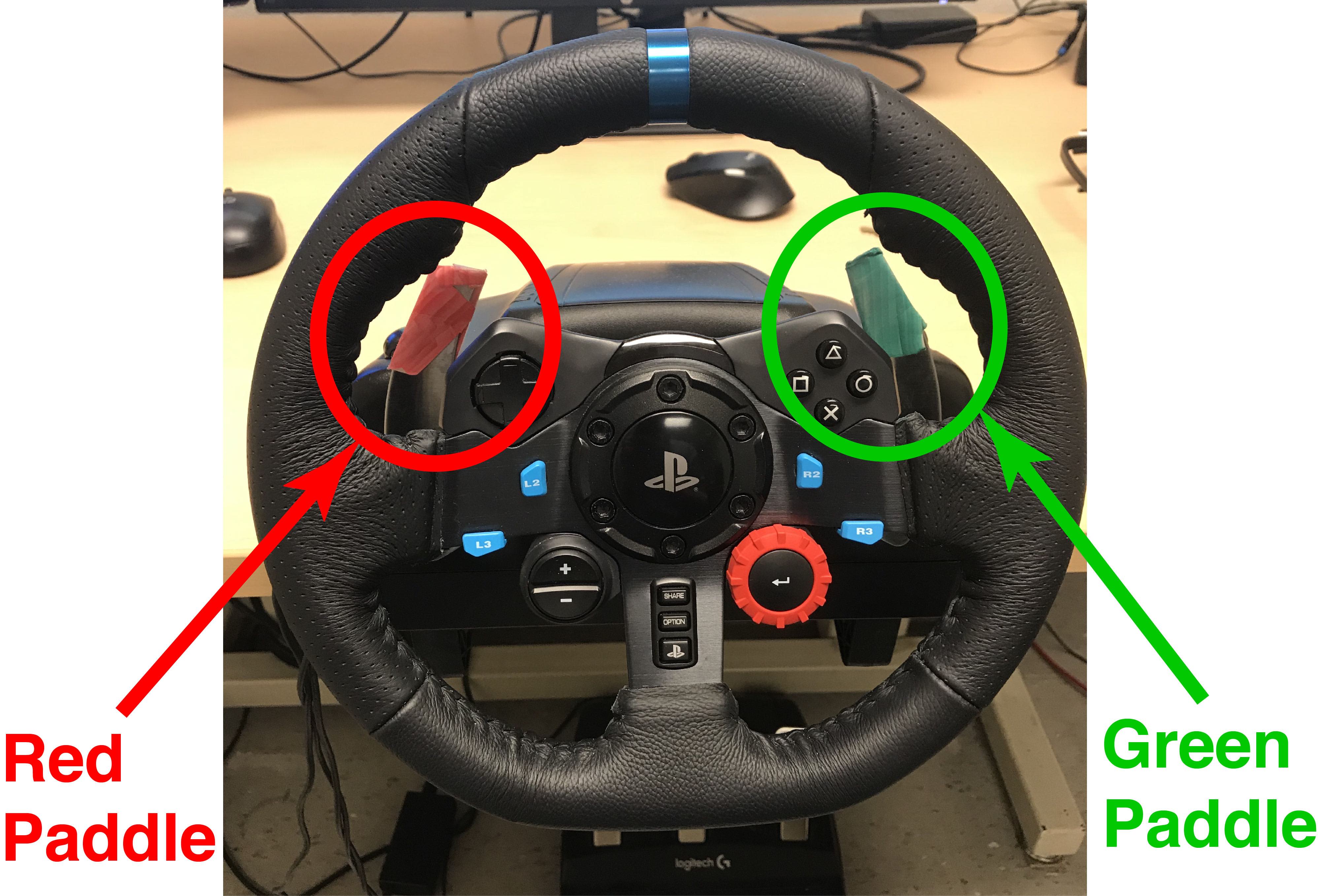}
    }
        \vspace{-2mm}
\caption{Dual-task shared control simulation platform}
\label{fig:test_bed}
\end{figure}

A dual-task shared control simulation platform was used in Experiment 1. Participants performed two tasks simultaneously, a driving task and a surveillance task as shown in Fig.~\ref{fig:test_bed}. 
% The screen for driving task is approximately 95 cm in front of the subject. The angles between the screen for driving task and the screen for surveillance task is 150 degrees. 

In the driving task, a participant and the autonomy shared the control of the HMMWV at a fixed speed of 15 mph, with the goal to complete a track with minimal deviation from the centerline. To emulate degraded localization due to sensor uncertainty, an offset was introduced such that the autonomy tracked a line which deviated from the centerline by 1 m. The non-adaptive shared control scheme was used in Experiment 1. The non-adaptive shared control scheme described in Sec. \ref{sec:nonAdaptiveSharedControl} was used in Experiment 1.

\begin{figure}[t]
    \centering
    \includegraphics[width = 0.7\linewidth]{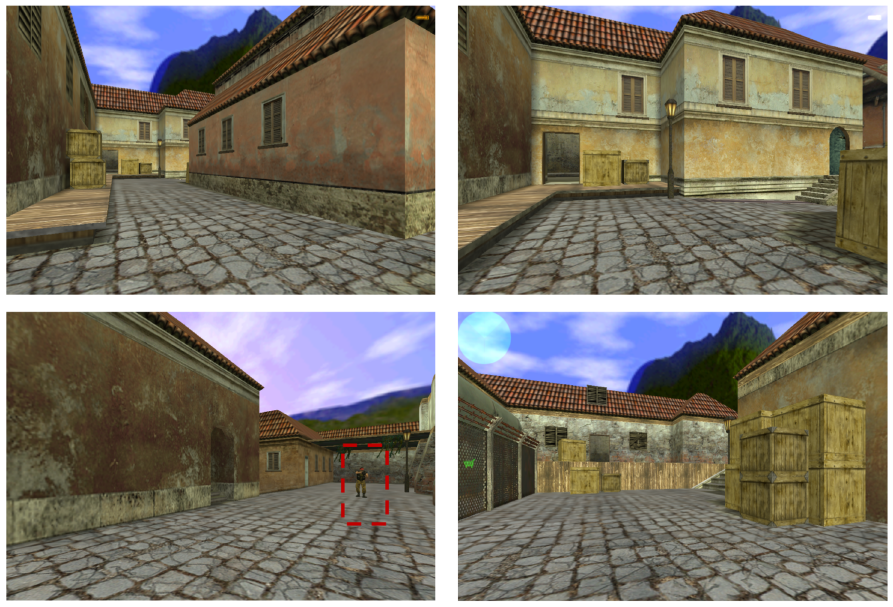}
        \vspace{-4mm}
    \caption{Illustration of the surveillance task. Lower left: threat.}
    %A threat is shown in the lower left photo.} 
    \label{fig:vsearch_images}
\end{figure}

In the surveillance task, the participant received image feeds and was asked to identify potential threats (Fig.~\ref{fig:vsearch_images}). If the participant identified a threat, s/he pressed the red paddle at the steering wheel to report ``danger". Otherwise, the participant pressed the green paddle to report ``clear" (Fig.~\ref{fig:testbed_pedal}).
%If the participant identified a threat, s/he reported "danger" by pressing the red paddle at the steering wheel. Otherwise, the participant reported "clear" by pressing the green paddle (Fig.~\ref{fig:testbed_pedal}). 
% As the steering wheel can only rotate from -90 degree to 90 degree, the participant would not need to cross their hands and could always keep the hands on the steering wheel. (The testbed setting)

% The potential threat will appear in only one of the four images. 
% The surveillance task utilized a fixed pace design: each set of four images will only appear for a fixed time $T$ with 1 second white screen between each set of four images. The fixed time $T$ can be controlled during the track and are used to control human's workload (See details in Appendix in pilot study 2).
Participants received a new set of four images at a fixed time interval, with a 1~s white screen in between, and were responsible for detecting potential threats as accurately as possible. The fixed time interval was varied to manipulate the workload level (See Appendix B for more details).

\subsubsection{Experimental design}
We manipulated the workload of the experimental tasks (the driving and the surveillance task) by varying the time interval of the surveillance task. During the experiment, the participants drove on 6 different tracks, each lasting for approximately 3 min. Every track was equally segmented into 3 portions and each portion had a different time interval for the surveillance task, $1.5$, or $2.5$, or $6.5$~s. The order of presentation for the 3 time intervals on each track is balanced by two $3 \times 3$ Latin squares.
% and the order of the rows of the Latin square is randomly determined. 
% The non-adaptive haptic shared control scheme was applied during this experiment.

\subsubsection{Measures}
Participants 
% performed both the driving task and the surveillance task on 6 different tracks. They 
wore a pair of the Tobii Pro Glasses 2 and their gaze points were recorded at 30Hz.

\subsubsection{Experimental procedure}
Participants provided a signed informed consent and filled in a demographic survey. After that, they received a training. Participants were first trained on the driving task alone, followed by the surveillance task alone. After that, they performed both the driving and the surveillance task. 
% on 3 different tracks. Each track had a different time intervals for the the surveillance task, $6.5$, $2.5$, and $1.5$ seconds.

% participants first performed a driving only task to get familiar with driving with the non-adaptive haptic shared control autonomy which takes approximately 1.5 minutes and then performed three trials of surveillance task with $6.5, 2.5, 1.5$ second fixed time where each trial takes approximately 60 seconds. 
% After that, the participants performed driving and surveillance task together on 3 different tracks with different surveillance task fixed time $6.5, 2.5, 1.5$ second. Each track takes approximately 1.5 minutes. The order of the surveillance task fixed time ($6.5, 2.5, 1.5$ second) is designed to help the participants to build capability to perform surveillance task with different difficulties gradually.

After the training session, participants were assisted to wear the eye tracker and underwent the calibration. With the normal room light and without any specific tasks, the experimenter measured each participant's baseline pupil diameter twice, each about 30~s. During the experiment, participants performed the driving task and the surveillance task on 6 different tracks, each lasting approximately 3 min. 
% Each track was equally segmented into 3 portions with 3 different surveillance task time intervals, $1.5$, $2.5$, and $6.5$ seconds. 
% After each track, the participants were asked to fill a post-survey about the workload and difficulty during each portion of the track.

% After finishing all the six trials, the subjects were required to fill a debriefing survey about outstanding questions and their opinions of or suggestions for the experiment they had just completed. 

\subsection{Results}
\subsubsection{Data processing}
Participants drove on 6 different tracks in this experiment. As mentioned above, each track was segmented into 3 portions and each portion had a different time interval for the surveillance task. The portion with the 1.5~s time interval was considered as the high workload portion, and the portion with the 6.5~s time interval the moderate workload portion. The ground truth labels were determined in two pilot studies (see Appendix A and B for details).
For each track, we randomly selected 5 sequences of data and each sequence lasted 4~s. 
% Therefore, for each participants, we obtained 8 sequences of data.  

% We treat the data from $6.5$ second fixed time in the surveillance task as high workload and $1.5$ second as moderate workload (see Appendix B for details).

% For each participant and each portion of each track, we randomly select 5 sequences of data with 4~s time window. We treat the data from $6.5$ second fixed time in the surveillance task as high workload and $1.5$ second as moderate workload (see Appendix B for details).

\subsubsection{Evaluation of the workload estimation performance}
Due to the small dataset of 12 participants, we used the holdout method \cite{kim2009estimating} for cross-validation and tested the performance of our proposed method. In each run of the holdout, we randomly selected data of 3 participants as the testing dataset and data of the remaining 9 participants as the training dataset. To find the best number of hidden states, we varied the number of hidden states from 2 to 10 for the HMM and ran 100 holdouts for each number of hidden states. The results indicate that 2 was the best number of hidden states.

We then ran another 100 holdouts to evaluate the performance of the HMM for workload estimation. Precision, recall and $F_1$ score were used as performance metrics, where $\text{precision} = \frac{\#\text{true positives}}{\#\text{true positives} + \#\text{false positives}}$ and $\text{recall} = \frac{\#\text{true positives}}{\#\text{true positives} + \#\text{false negatives}}$. For our multi-classification problem, the precision is the mean precision of all classes and the recall is the mean recall of all classes. $F_1 = 2\frac{\text{precision}\cdot \text{recall}}{(\text{precision}+\text{recall})}$. Table~\ref{tab:HMM_performance} shows the mean and standard error of each performance metric. The results show that the HMM model achieved a 0.66 $F_1$ score, 0.67 precision and 0.66 recall.

% In the Experiment 2, we used HMM learned with the data from the all 12 participants to estimate the participant's workload in real time.

\begin{table}[t]
\centering
\caption{Performance of the HMM}
    \vspace{-3mm}

\label{tab:HMM_performance}
\begin{tabular}{c|cccc}
\hline
    & $F_1$               & Precision           & Recall                         \\ \hline
HMM & $0.664 \pm 0.005$ & $0.668 \pm 0.005$ & $0.660 \pm 0.005$  \\ \hline
\end{tabular}
\end{table}

\section{EXPERIMENT 2}
\subsection{Introduction}
% In Experiment 1, we see that the HMM model can achieve a 0.66 $F_1$ score, 0.67 precision and 0.66 recall. 
In Experiment 2, we tested two different haptic shared control schemes: the adaptive haptic shared control and non-adaptive haptic shared control schemes. The adaptive haptic shared control scheme adapted to the estimated real-time workload, and the participant's eyes on road and torque input. We used the HMM learned with the data from all the 12 participants to estimate the participant's workload in real time.
\subsection{Method}
% This study complied with the American Psychological Association code of ethics and was approved by the Institutional Review Board at the University of Michigan.

\subsubsection{Participants}
A total of 13 students participated in the experiment. Data of 1 participant were discarded due to the wrong experiment setup. The remaining 12 participants were on average 22.3 years old (\textit{SD} = 3.7 years) and had an average of 5.7 years of driving experience (\textit{SD} = 3.9 years). All participants had normal or corrected-to-normal vision.
% and their eyes are able to be calibrated by the eye tracker.

\subsubsection{Apparatus and stimuli}
The same dual-task shared control simulation platform was used in this experiment as in Experiment 1. Both the adaptive haptic shared control and the non-adaptive haptic shared control were used in this experiment. 

\subsubsection{Experimental design}
The experiment used a within-subjects design with two independent variables. The first independent variable was the haptic shared control scheme (adaptive haptic shared control vs. non-adaptive haptic shared control). The second independent variable was the the secondary task urgency (1.5~s vs. 6.5~s). Each participant experienced four tracks in the experiment. On each track, one type of haptic shared control scheme was used. In addition, each track was segmented into two portions, one portion with high urgency secondary task (1.5~s) and the other with low urgency secondary task (6.5~s). The resulting four test conditions are shown in Table \ref{tab:test_case}. 
% We designed 4 different cases combining different surveillance task fixed time on each portion of the track and different haptic shared control schemes as shown in Table~\ref{tab:test_case}. 
The presentation of test conditions followed a $4 \times 4$ Latin square design to eliminate potential order effects.

% \begin{table}[]
% \caption{Different Test Conditions}
% \label{tab:test_case}
% \begin{tabular}{c|ccc}
% \hline
% Test Condition & First half of the track   & Second half of the track   & Haptic Shared Control Scheme \\ \hline
% 1         & 1.5 seconds & 6.5 seconds & Non-adaptive                 \\ \hline
% 2         & 1.5 seconds & 6.5 seconds & Adaptive                     \\ \hline
% 3         & 6.5 seconds & 1.5 seconds & Non-adaptive                 \\ \hline
% 4         & 6.5 seconds & 1.5 seconds & Adaptive                     \\ \hline
% \end{tabular}
% \end{table}

% \usepackage{multirow}
\begin{table}[t]
\centering
 \caption{Four Test Conditions}
     \vspace{-3mm}
 \label{tab:test_case}
\begin{adjustbox}{width=0.9\linewidth}
\begin{tabular}{cccc}
\hline
\multirow{2}{*}{Condition} & \multicolumn{2}{c}{Surveillance task urgency} & \multirow{2}{*}{Haptic Shared Control Scheme} \\
  & First half of the track & Second half of the track &  \\ \hline
1 & 1.5~s & 6.5~s & Non-adaptive \\
2 & 1.5~s & 6.5~s & Adaptive \\
3 & 6.5~s & 1.5~s & Non-adaptive \\
4 & 6.5~s & 1.5~s & Adaptive \\ \hline
\end{tabular}
 \end{adjustbox}
\end{table}

% \textit{\textbf{Independent variables.}} The independent variable in the experiments was {\color{red}{the haptic shared control scheme}}: adaptive haptic shared control and non-adaptive haptic shared control.

\subsubsection{Measures}
Five dependent variables were collected in the experiment: participants' self-reported workload and trust in the shared control autonomy,  participants' control effort, driving task performance and surveillance task performance. After each track, participants reported their workload and trust for the first and second half of the track using two uni-dimensional scales. % (see Appendix C). 
The NASA TLX survey \cite{HART1988} and the Moray's trust survey \cite{MORAY:1996it} were presented to the participants such that they understood the meaning of workload and trust. 
% gain as a reference for different dimensions of workload and trust to report the overall workload and trust for each portion of the track. 
Participants' control effort was calculated as the average torque that a participant applied on the steeling wheel. %The measurement was acquired at the frequency of 100~Hz.
Driving task performance was evaluated by lane keeping error. The lane keeping error is calculated as the mean of the absolute deviation of the vehicle's position from the centerline. %The measurement was acquired at the frequency of 100~Hz. 
% and control effort which are participants' lane keeping error and torque applied, respectively
%The torque is obtained from a steering wheel model. Both measurements are
% \textcolor{red}{Jessie: @Yifan, here can you describe how the lane keeping error was calculated? root mean square? We need to mention the sampling rate as well, for both the lane keeping error and torque}
The surveillance task performance was measured using the detection accuracy.

\subsubsection{Experimental procedure}
Participants provided a signed informed consent and filled in a demographic survey. After that they were assisted to wear the eye tracker with calibration. With the normal room light and without any specific tasks, the experimenter measured each participants' baseline pupil diameter twice each about 30~s before the training.

During the training session, the participants first performed two trials of driving task only, one with the non-adaptive haptic shared control and one with the adaptive haptic shared control. Each trial took approximately 1.5~min. Then the participants performed three trials of the surveillance task only. Each trial took approximately 60~s. After that, the participants performed four trials of the combined driving and surveillance task.
% together on 4 different tracks.  with different surveillance task fixed time and different haptic shared control schemes with the following orders: 1. $6.5$ seconds with non-adaptive haptic shared control; 2. $6.5$ seconds with adaptive haptic shared control; 3. $1.5$ seconds with non-adaptive haptic shared control; 4. $1.5$ seconds with adaptive haptic shared control. Each track takes approximately 1.5 minutes. The order of the surveillance task fixed time is designed to help the participants to build capability to perform surveillance task with different difficulties gradually.

During the official experiment, participants performed the driving task and the surveillance task on 4 different tracks with different test cases as described in Table~\ref{tab:test_case}. Each trial took approximately 3 min. After each trial, the participants were asked to fill a post survey about the workload and trust during each portion of the track.

\subsection{Experiment 2 Results}

 Two-way repeated measures Analysis of Variance (ANOVAs) were conducted with the shared control scheme and the surveillance task urgency as the within-subjects variables.
%  Sidak adjustments \cite{Abdi2007} were employed for pairwise comparisons. Effect size was calculated as Eta squared $\eta_{p}^{2}$ for ANOVAs and as Cohen's $d$ for pairwise \cite{Field:2013:DSU:2502692}. 
  Results are reported as significant for $\alpha < .05$. 
%   \cite{Gelman:2013eq}. 
%  Effect size is considered as small if $d = .2$, medium if $d = .5$, and large if $d = .8$ \cite{Cohen:1992bp}.
Table~\ref{tab:raw_data} summarizes the mean and standard error (SE) values of the participants' self-reported workload and trust as well as driving task performance, surveillance task performance and their exerted torque.
% Due to device malfunction, some data points were missed. The resulting sample size \textit{N} for each dependent variable is detailed in Table 1.

% \usepackage{multirow}
\begin{table}[t]
\centering
\caption{Mean and Standard Error (SE) of workload, trust, lane keeping error, detection accuracy and torque }
    \vspace{-3mm}
\label{tab:raw_data}
\begin{adjustbox}{width=0.9\linewidth}
\begin{tabular}{cccccc}
\hline
\multirow{3}{*}{Metrics} & \multirow{3}{*}{N} & \multicolumn{4}{c}{Surveillance task urgency}                                           \\ \cline{3-6} 
                         &                    & \multicolumn{2}{c}{$1.5$ s}           & \multicolumn{2}{c}{$6.5$ s}         \\ \cline{3-6} 
                         &                    & Adaptive              & Non-adaptive        & Adaptive            & Non-adaptive        \\ \hline
Workload                 & 12                 & $13.96 \pm 0.82$ & $14.08\pm0.87$ & $7.83\pm0.81$  & $8.71\pm0.97$  \\ \hline
Trust                    & 12                 & $4.04\pm0.37$    & $3.63\pm0.30$  & $3.92\pm0.32$  & $3.29\pm0.38$  \\ \hline
Lane keeping error (m)   & 12                 & $0.28\pm0.033$    & $0.36\pm0.045$  & $0.21\pm0.03$  & $0.26\pm0.04$  \\ \hline
Detection accuracy (\%)  & 12                 & $93.43\pm1.38$   & $91.86\pm1.13$ & $94.30\pm1.77$ & $96.54\pm1.18$ \\ \hline
Torque (Nm)              & 12                 & $0.36\pm0.03$    & $0.73\pm0.03$  & $0.30\pm0.02$  & $0.79\pm0.01$  \\ \hline
%Assistance Level $\beta$              & 12                 & $1.03\pm0.04$    & $1\pm0$  & $0.82\pm0.04$  & $1\pm0$  \\ \hline

\end{tabular}
\end{adjustbox}
\end{table}

% % Please add the following required packages to your document preamble:
% % \usepackage{multirow}
% \begin{table}[]
% \begin{tabular}{cccccc}
% \hline
% \multirow{3}{*}{Metrics} & \multirow{3}{*}{N} & \multicolumn{4}{c}{Surveillance task urgency}                                           \\ \cline{3-6} 
%                          &                    & \multicolumn{2}{c}{$6.5$ seconds}         & \multicolumn{2}{c}{$1.5$ seconds}           \\ \cline{3-6} 
%                          &                    & Non-adaptive        & Adaptive            & Non-adaptive        & Adaptive              \\ \hline
% Lane keeping error (m)   & 12                 & $0.2571\pm0.03981$  & $0.2064\pm0.02674$  & $0.3586\pm0.04498$  & $0.2750\pm0.03253$    \\ \hline
% Detection accuracy (\%)  & 12                 & $96.5430\pm1.18234$ & $94.2995\pm1.77048$ & $91.8559\pm1.12530$ & $93.4341\pm1.37720$   \\ \hline
% Torque (Nm)              & 12                 & $0.7926\pm0.01407$  & $0.3014\pm0.02031$  & $0.7310\pm0.03481$  & $0.3639\pm0.03036$    \\ \hline
% Workload                 & 12                 & $8.7083\pm0.97012$  & $7.8333\pm0.81029$  & $14.0833\pm0.86785$ & $16.4583 \pm 2.45525$ \\ \hline
% Trust                    & 12                 & $3.2917\pm0.37668$  & $3.9167\pm0.31881$  & $3.6250\pm0.29596$  & $4.0417\pm0.37162$    \\ \hline
% \end{tabular}
% \end{table}

\subsubsection{Participants' Workload}
Both control scheme and surveillance task urgency influence participants' self-reported workload. 
With the adaptive shared control, participants reported lower workload ($F(1,11) = 5.18$, $p =.044$). When the surveillance task was less urgent, participants reported lower workload ($F(1,11) = 20.26$, $p <.001$). 
\subsubsection{Trust in Automation}
Participants trusted the shared control autonomy more when the autonomy was adaptive ($F(1,11) = 12.76$, $p =.004$). The effect of surveillance task urgency on trust was not significant.

% With adaptive shared control, participants exerted significantly less control effort. In addition, results revealed a significant interaction effect between control scheme and surveillance task urgency ($F(1,11) = 11.42$, $p =.006$). When the surveillance task was less urgent (6.5 seconds), the adaptive shared control scheme led to a larger drop in torque. 

\begin{figure}[t]
\centering\subfloat[Workload]{\includegraphics[width = 0.45\linewidth]{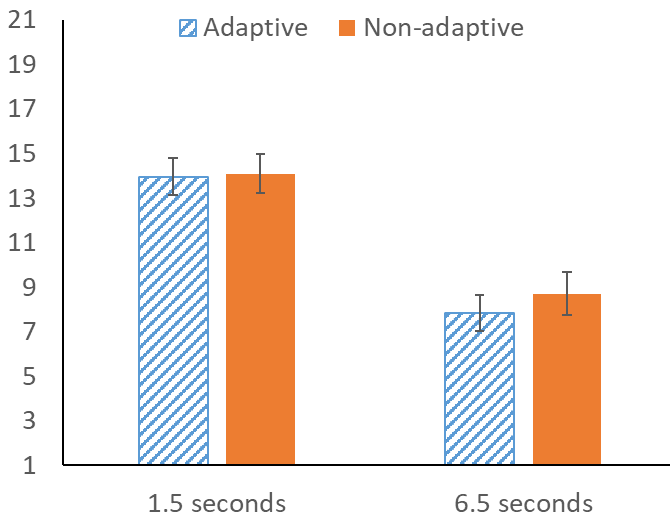}\label{fig:exp2_workload}}
\subfloat[Trust]{\includegraphics[width = 0.45\linewidth]	{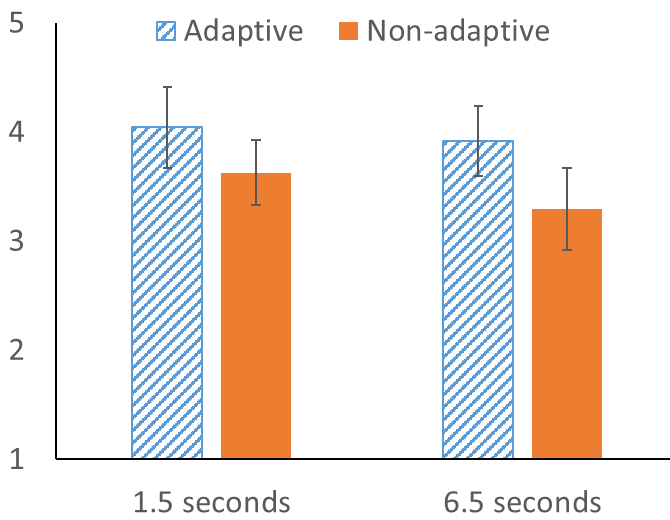}\label{fig:exp2_trust}}
    \vspace{-2mm}
  \caption{Mean and standard error (SE) values of self-reported workload and trust}
  \label{fig:exp2_workload_and_trust}
\end{figure}

% \begin{figure}[h]
%     \centering
%     \includegraphics[width = 0.8\linewidth]{Figures/preceived_workload.png}
%     \caption{} 
%     \label{fig:exp2_workload}
% \end{figure}

% \begin{figure}[h]
%     \centering
%     \includegraphics[width = 0.8\linewidth]{Figures/trust.png}
%     \caption{} 
%     \label{fig:exp2_trust}
% \end{figure}

\subsubsection{Driving Task Performance}
\begin{figure}[t]
    \centering
    \subfloat[Lane keeping error (m)]{\includegraphics[width = 0.45\linewidth]{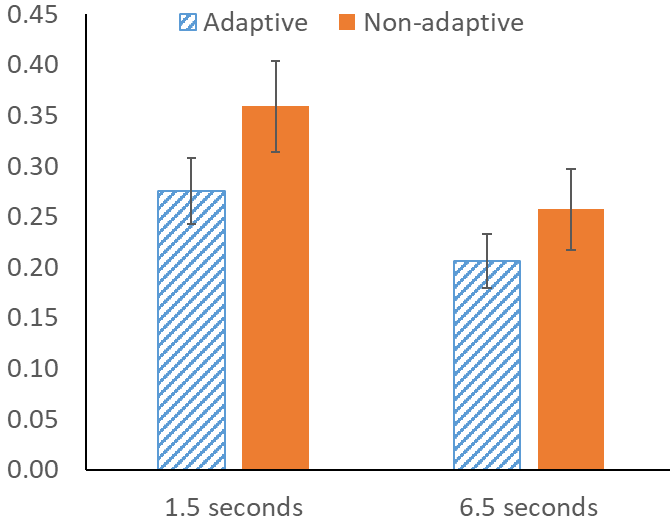}\label{fig:exp2_lane_keeping_error}}
    \subfloat[Detection accuracy (\%)]{\includegraphics[width = 0.45\linewidth]{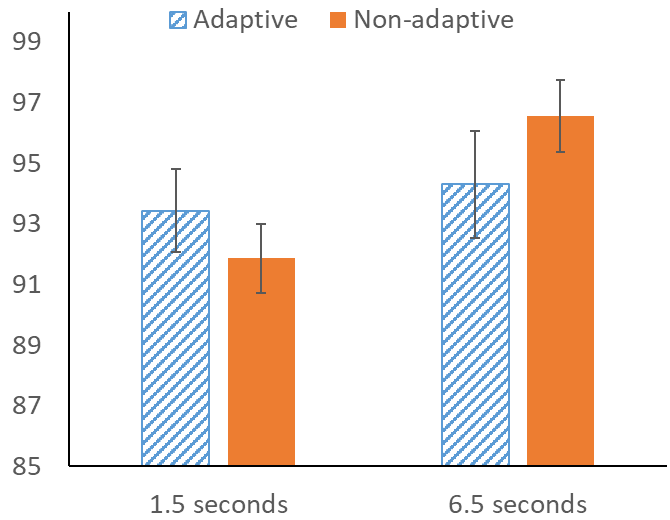}\label{fig:exp2_detection_accuracy}}
        \vspace{-2mm}
    \caption{Mean and standard error (SE) values of lane keeping error (m) and surveillance task detection accuracy (\%).} 
\end{figure}

Results revealed that the haptic shared control scheme and the surveillance task urgency significantly affected the driving task performance. Participants had smaller lane keeping errors when using the adaptive shared control autonomy ($F(1,11) = 7.593$, $p = .019$), and when the surveillance task was less urgent ($F(1,11) = 96.33$, $p < 0.001$) (Fig. \ref{fig:exp2_lane_keeping_error}). There was also an interactive effect between the control scheme and surveillance task urgency ($F(1,11) = 6.141$, $p =.031$). Using adaptive shared control led to a large reduction in lane keeping error when the surveillance task was more urgent. 
% A simple effect analysis revealed a significant difference between adaptive and non-adaptive shared control when the surveillance task was more urgent ($F(1,11) = 10.997$, $p = .007$), and a non-significant difference when the surveillance task was less urgent.

\subsubsection{Surveillance Task Performance}
For the surveillance task, task urgency significantly influenced the detection accuracy  ($F(1,11) = 6.73$, $p =.025$). Detection accuracy was higher when the task was less urgent. The effect of the shared control scheme was non-significant (Fig. \ref{fig:exp2_detection_accuracy}).

\subsubsection{Participants' Control Effort}
There was a significant effect of shared control scheme on participants' control effort ($F(1,11) = 217.66$, $p <.001$). With adaptive shared control, participants exerted significantly less control effort. The effect of surveillance task urgency on participants' control effort was non-significant. In addition, results revealed a significant interaction effect between control scheme and surveillance task urgency ($F(1,11) = 11.42$, $p =.006$). When the surveillance task was less urgent (6.5~s), the adaptive shared control scheme led to a larger drop in torque.

\begin{figure}[t]
    \centering
    \includegraphics[width = 0.5\linewidth]{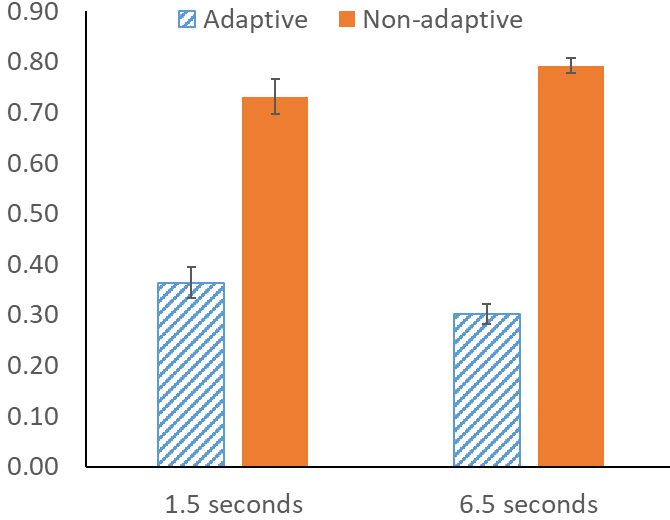}
    \vspace{-3mm}
    \caption{Mean and standard error (SE) values of participants' torque (Nm)} 
    \label{fig:exp2_torque}
\end{figure}

\subsection{Experiment 2 Discussion}
% When the scheme is designed, we consider two design principles. On one hand, when the subject experiences 
% moderate workload, since he/she focuses on the driving task, we remove the unnecessary fight with autonomy by decreasing assistance level and hence expect a drop in human torque. On the other hand, we increase assistance level when subject has a high workload and focuses on the surveillance task, expecting an decrease in lane keeping error.
\subsubsection{Participants' Workload}
Participants' self-reported workload decreased when using the adaptive shared control scheme and when the surveillance task became less urgent. The results could have resulted from the following reasons. First, the 6.5~s surveillance task urgency imposed a smaller temporal demand on participants. Second, the participants' control effort was smaller with the adaptive control scheme. Third, participants' driving task performance was higher with the adaptive control scheme and when the surveillance task was less urgent. 

% Even if with adaptive haptic shared control scheme both the lane keeping error and the human torque decreased, the participants still need effort to perform the driving and surveillance task, thus the effect of haptic shared control scheme on workload was not significant. 

% {\color{red}{Not sure how to explain why the trust is significant.}}
\subsubsection{Trust in Automation}
Our result is consistent with prior research that human operators' trust in automation is determined by the autonomy's performance \cite{yang2017evaluating,Du2019}. Human operators perceived both the driving and the surveillance task performance continuously, based on which they adjusted their trust in automation. As the driving task performance increased with the adaptive control scheme, trust increased accordingly. %\Yifan{Vishnu: the autonomy will not be perfect (otherwise we would hand over control completely), and so by blending intelligently between the human and the autonomy, we get the best of both worlds}

\subsubsection{Driving Task Performance}
The results showed that the adaptive shared control scheme benefited the driving task performance, especially when participants were under a high workload. Based on the design of the adaptive haptic shared control scheme, with the same input torque, when the human operator has a high workload and focuses on the surveillance task, the assistance level is increased. %(average $\beta = 1.03$ with adaptive shared control scheme compared with $\beta = 1$ with non-adaptive shared control scheme when surveillance task is more urgent). 
The increment in the assistance level is expected to aid the driving task and reduce the lane keeping error. This design principle was supported by the experimental results.
\subsubsection{Surveillance Task Performance}
 As the surveillance task became more urgent and more demanding, the surveillance task performance decreased significantly. This result is consistent with prior research that when workload increased from moderate to high level, task performance would decrease \cite{lu2019workload}.  
\subsubsection{Participants' Control Effort}
Our results indicate that with adaptive shared control participants exerted significantly less amount of control effort in both low and high workload conditions. The results can be explained as follows: First, as the participants' trust toward the adaptive shared control scheme is significantly higher than the non-adaptive control scheme, participants had a higher tendency to yield to the autonomy, resulting in smaller input torque. Second, according to the design of the adaptive shared control scheme, with the same input torque, when the human operator experiences moderate workload and focuses on the driving task,  the assistance level is reduced.% (average $\beta = 0.82$ with adaptive shared control scheme compared with $\beta = 1$ with non-adaptive shared control scheme when surveillance task is less urgent)
With a reduced assistance level, regardless of whether the human yields to or fights with the autonomy, the human operator's torque is expected to be smaller.
% Therefore, we would expect a drop in the human operator's torque. This principle is also supported by the result -- a significant difference on the human's torque between the two haptic shared control schemes. As the workload was mostly estimated to be low during this condition, the assistant level ($\beta < 1$) in the adaptive haptic shared control scheme was smaller than the assistant level in the non-adaptive haptic shared control ($\beta = 1$), resulting in smaller torque from autonomy in the adaptive haptic shared control. 

% However, as the torque from autonomy with adaptive shared control in 1.5-second surveillance task urgency is larger than in 6.5-second surveillance task urgency, the human's torque required to fight with autonomy is larger in 1.5-second surveillance task urgency than in 6.5-second surveillance task urgency in case human did not yield the autonomy. Thus, the shared control scheme has more effect on human's torque in 6.5-second surveillance task urgency than in 1.5-second surveillance task urgency.

% \section{5. GENERAL DISCUSSION}
\section{GENERAL DISCUSSION and CONCLUSION}
In this study, we proposed an adaptive haptic shared control scheme by designing a heuristic function for assistance level considering human's workload, torque and eyes on road. The results indicate that our  adaptive haptic shared control scheme leads to lower self-reported workload, higher trust in automation, lower lane keeping error and lower human control effort.
To our best knowledge, this is the first study in which the human operator's workload was estimated in real-time and used as an input to an adaptive haptic shared control scheme. 

The findings should be viewed in light of the following limitations. First, a group-level workload estimation model was developed in the study, ignoring potential individual differences. 
% However, the variety of people is really large, one simple model learned from a small dataset does not fit all people. Also, people have different capabilities in completing the dual tasks, thus their true workload may be different even with the same surveillance task urgency. 
In this study, we assumed that participants experienced high workload when the surveillance task was more urgent and moderate workload when surveillance task was less urgent. This assumption may not hold for different individuals. In the future work, we will develop a personalized workload estimation model to account for individual differences. Second, the assistance level adaptation function is manually designed based on heuristics. In the future work, we would investigate other methods to find the optimal assistance level adaptation function.

%Third, the experiment only tested the effect of the adaptive haptic shared control scheme, but did not study each factor considered in the assistance level adaptation function: workload, torque and eyes on road. Further studies are required to see the effect of each single factor.

%Third, the experiment design only takes part of the possible conditions into account, such as high workload with low eyes on road rate. Conditions like high workload with high eyes on road cannot happen in our experiment design. In the future work, we would like to conduct more experiments to test other different scenarios.

% \newpage
\appendices
    \vspace{-3mm}
\section{Pilot Study 1 -- Track Selection}
% In this pilot study, we aimed to evaluate different tracks and select the tracks with similar difficulties and the difficulty on the track are uniformly distributed.

% In the experiment, we manipulated the difficulty of the 

In pilot study 1, we developed and selected 6 driving tracks with two considerations. First, the driving tracks should have the same difficulty. Second, along each track, the difficulty at every point should be roughly the same. The two considerations ensure that the difficulty of the dual-task mission can be easily manipulated by varying the surveillance task urgency, because the difficulty of the driving task is fairly constant.

\subsubsection{Participants}
10 participants (Age: Mean = 21.8 years, \textit{SD} = 2.7 years) took part in pilot study 1. All participants had normal or corrected-to-normal vision and hearing, with an average of 4.1 years of driving experience (\textit{SD} = 1.7 years). 

\subsubsection{Apparatus and stimuli}
Pilot study 1 used the same driving simulator as in Experiment 1 with driving task only and non-adaptive haptic shared control scheme.% Participants only performed the driving task, with the non-adaptive haptic shared control scheme. 
% Only the driving task was performed in this pilot study and no surveillance task.

\subsubsection{Experimental design}
\begin{figure}[h]
\centering
    \vspace{-8mm}

\subfloat[Track 1]{\includegraphics[width = 0.2\linewidth]{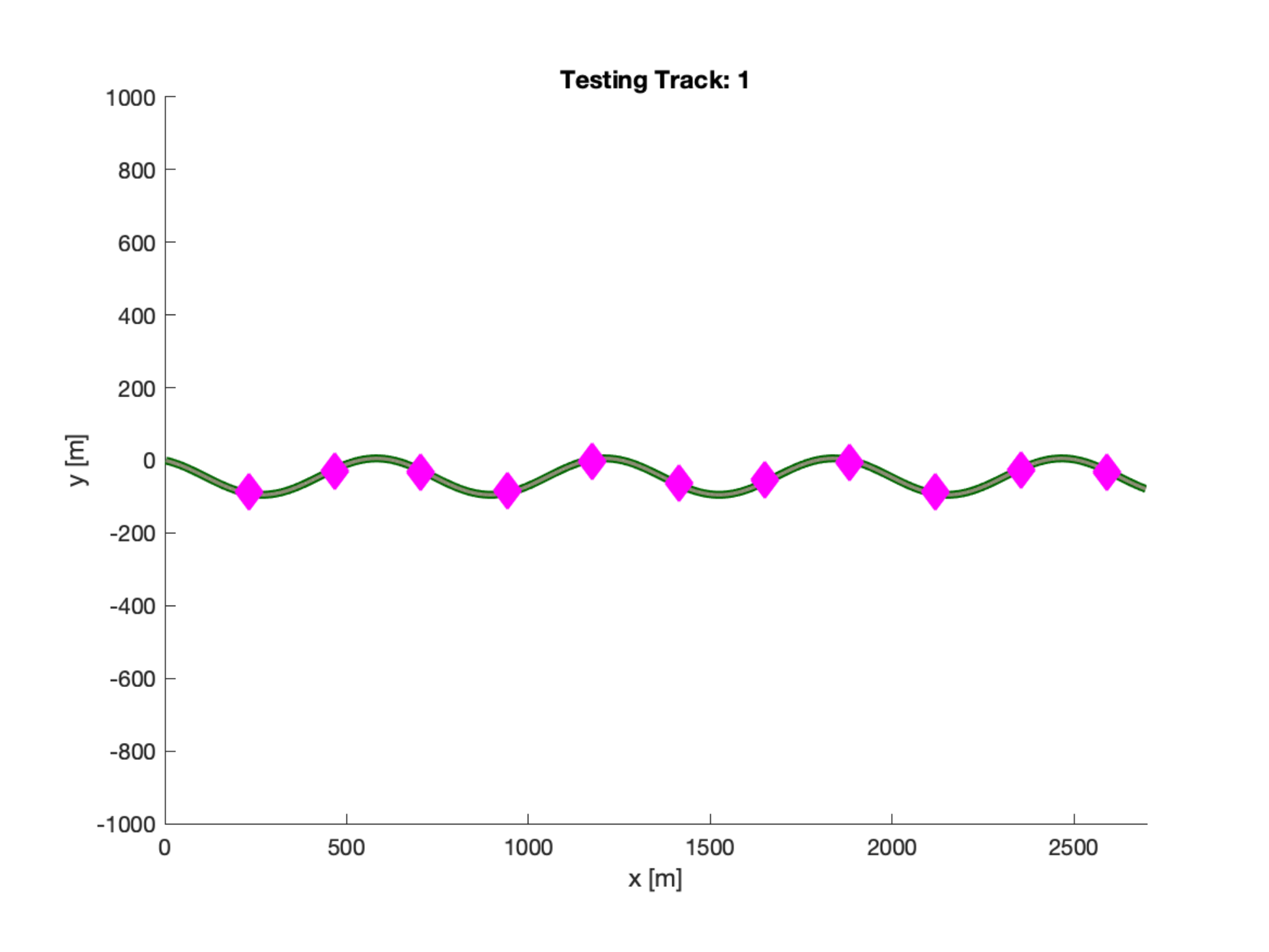}}
\subfloat[Track 2]{\includegraphics[width = 0.2\linewidth]	{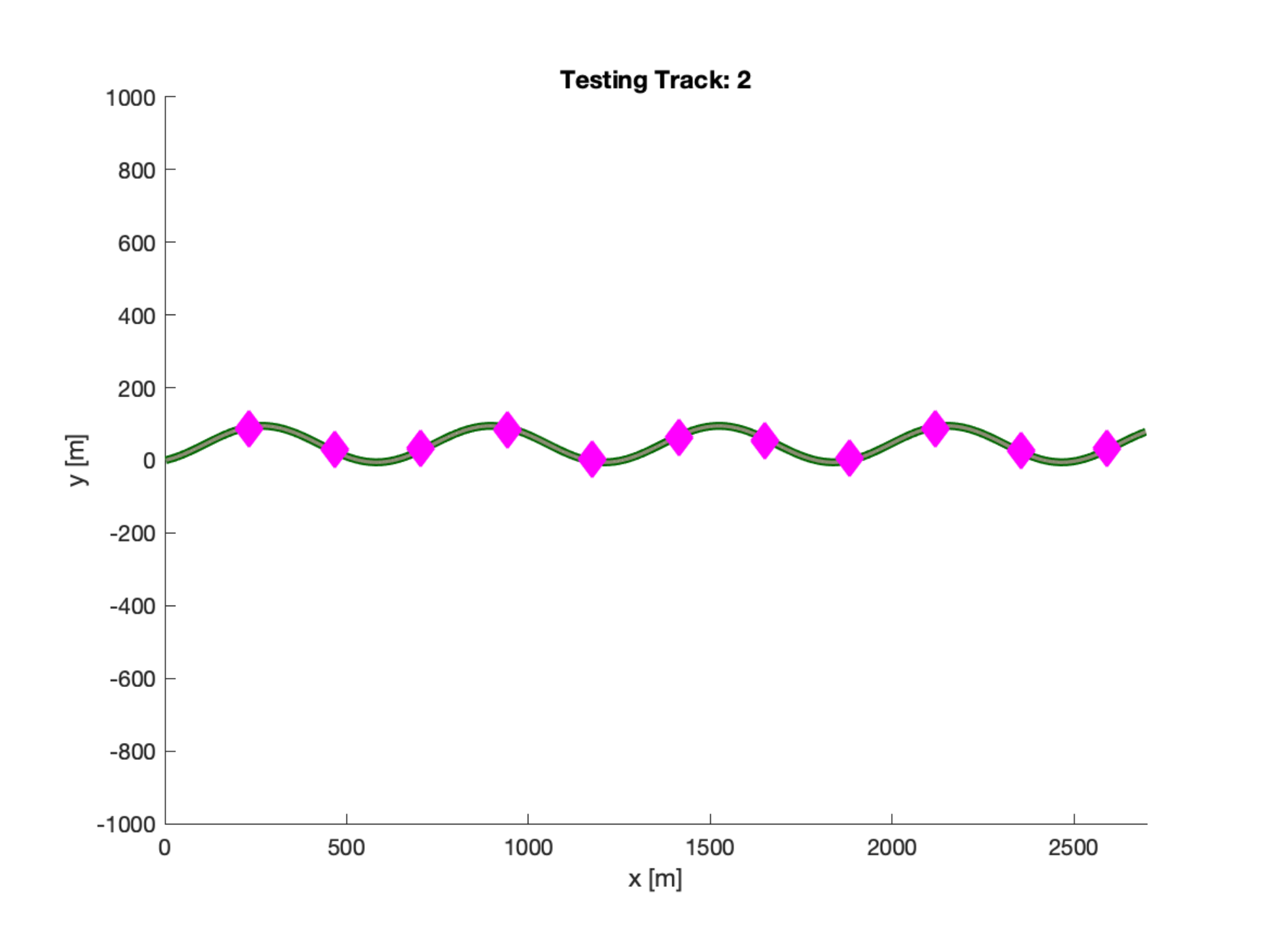}}
\subfloat[Track 3]{\includegraphics[width = 0.2\linewidth]	{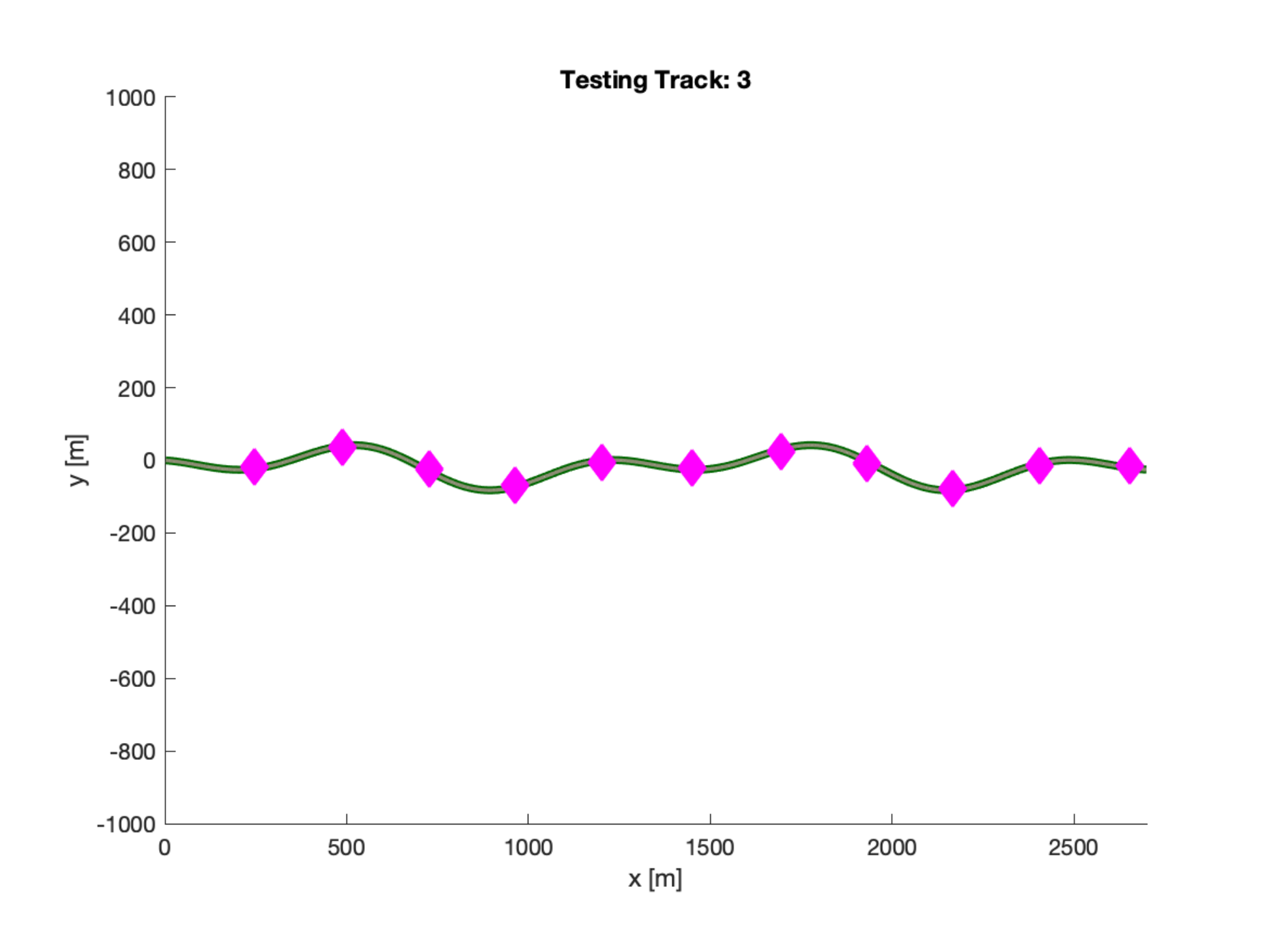}}
\subfloat[Track 4]{\includegraphics[width = 0.2\linewidth]{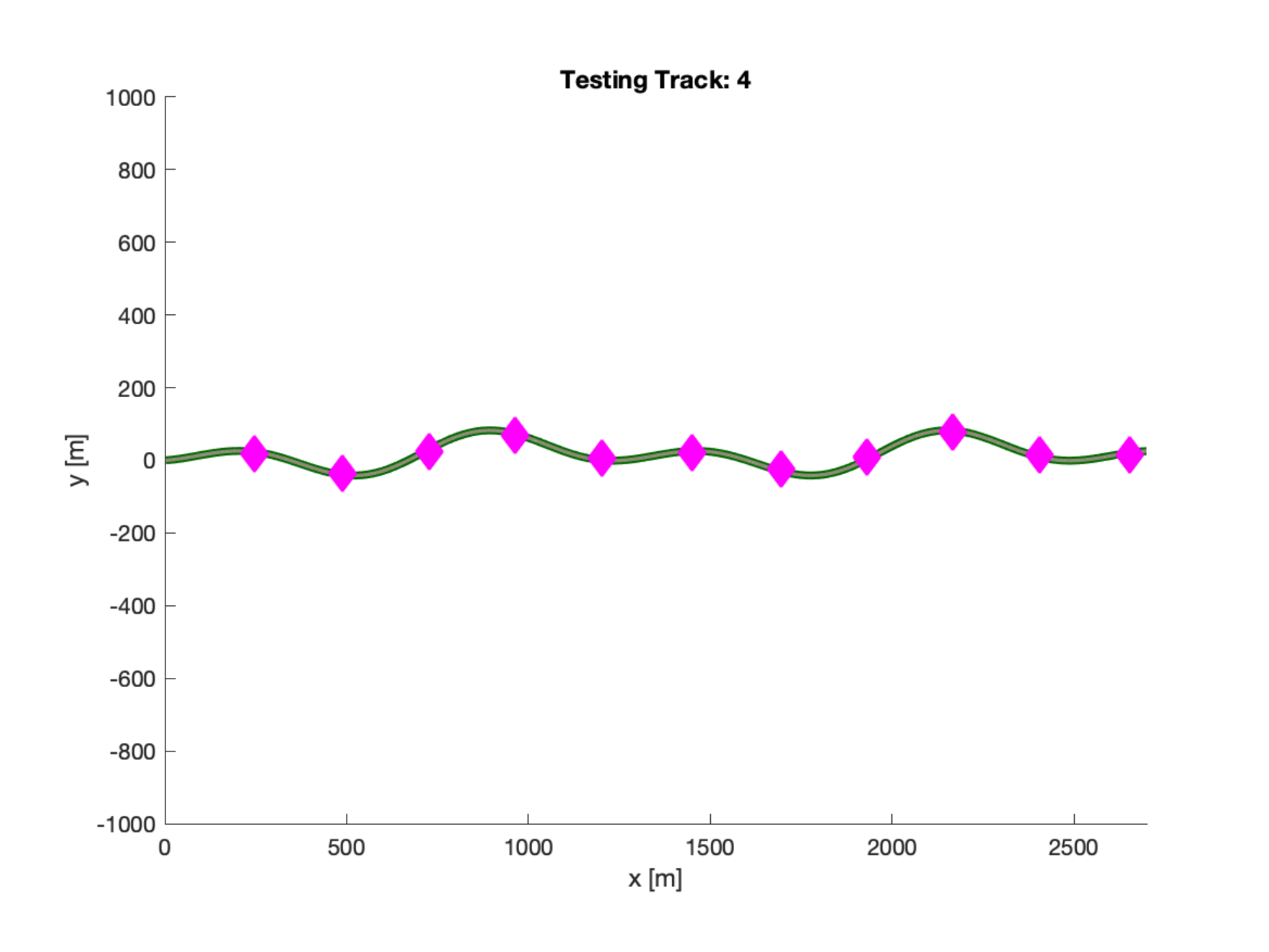}}
\subfloat[Track 5]{\includegraphics[width = 0.2\linewidth]	{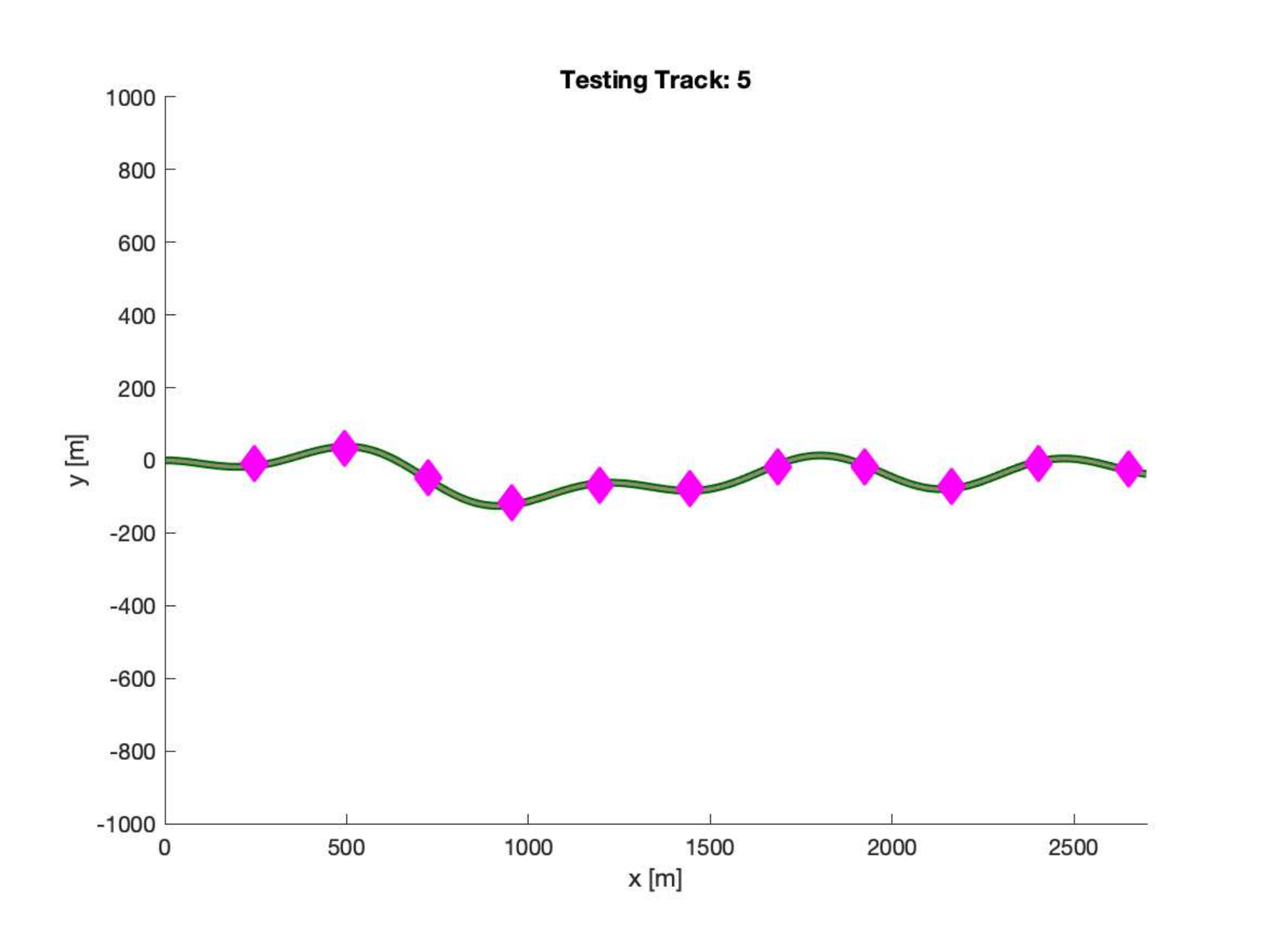}}

\centering
    \vspace{-3mm}

\subfloat[Track 6]{\includegraphics[width = 0.2\linewidth]	{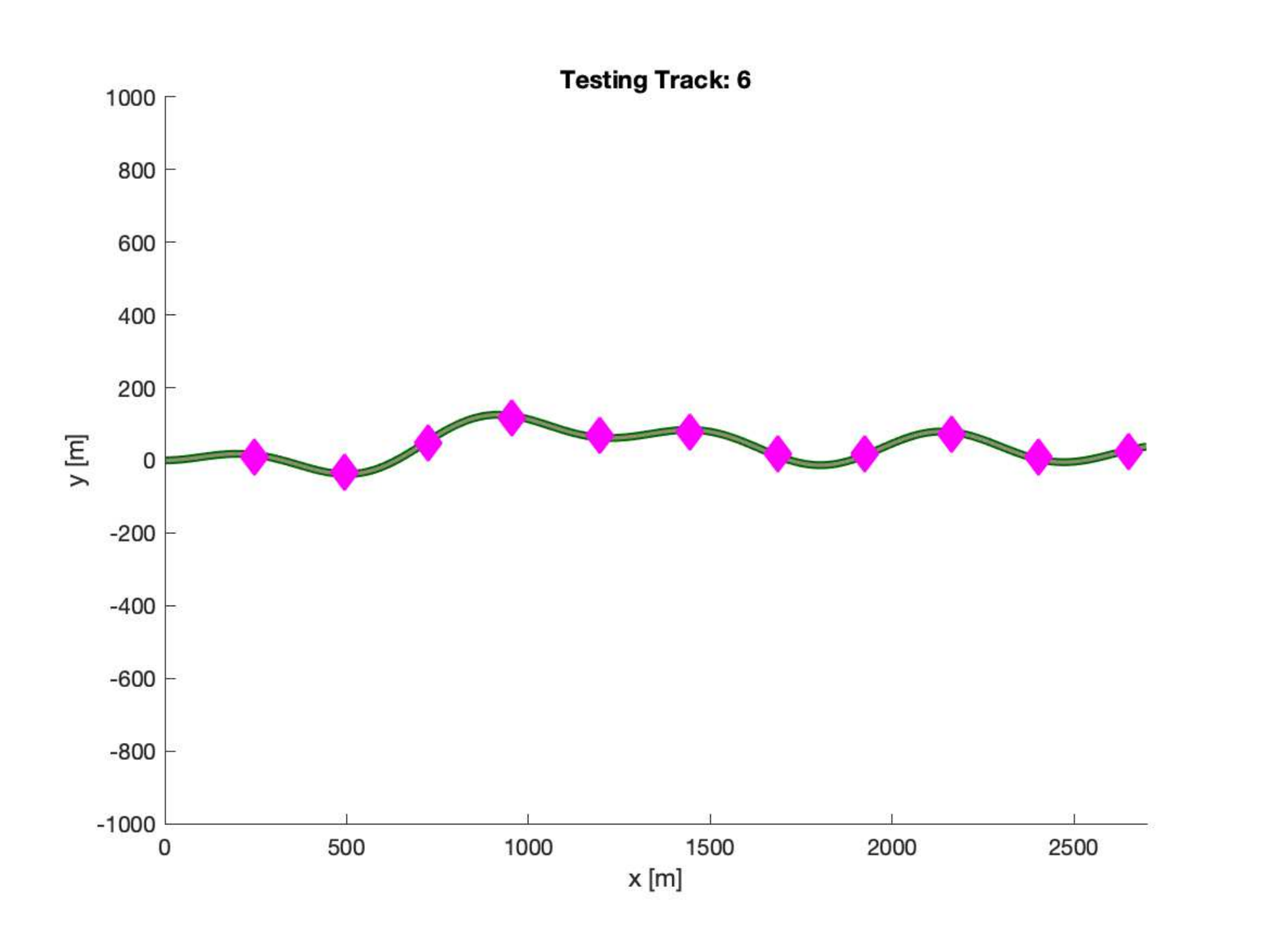}}
\subfloat[Track 7]{\includegraphics[width = 0.2\linewidth]	{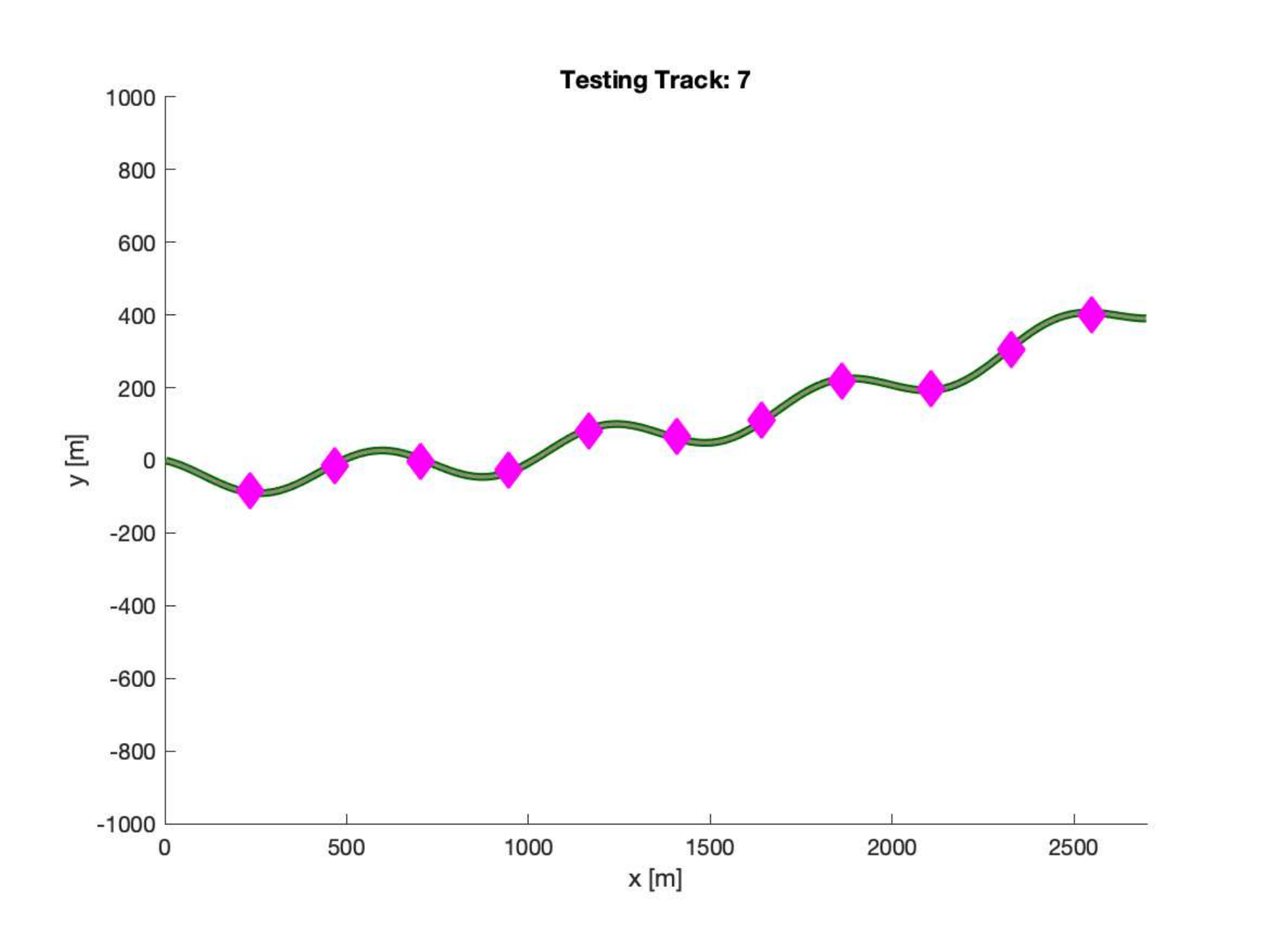}}
\subfloat[Track 8]{\includegraphics[width = 0.2\linewidth]	{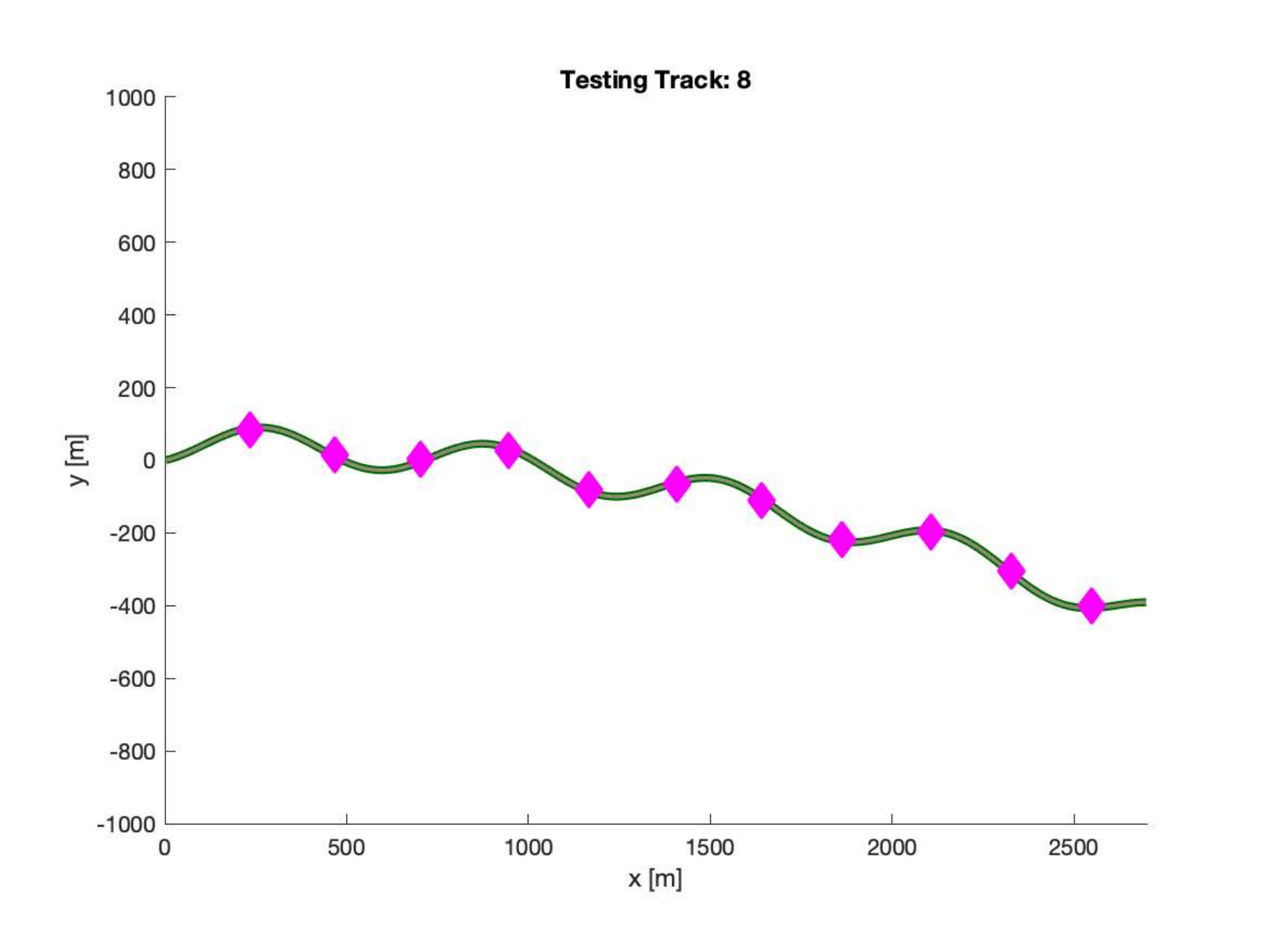}}
\subfloat[Track 9]{\includegraphics[width = 0.2\linewidth]{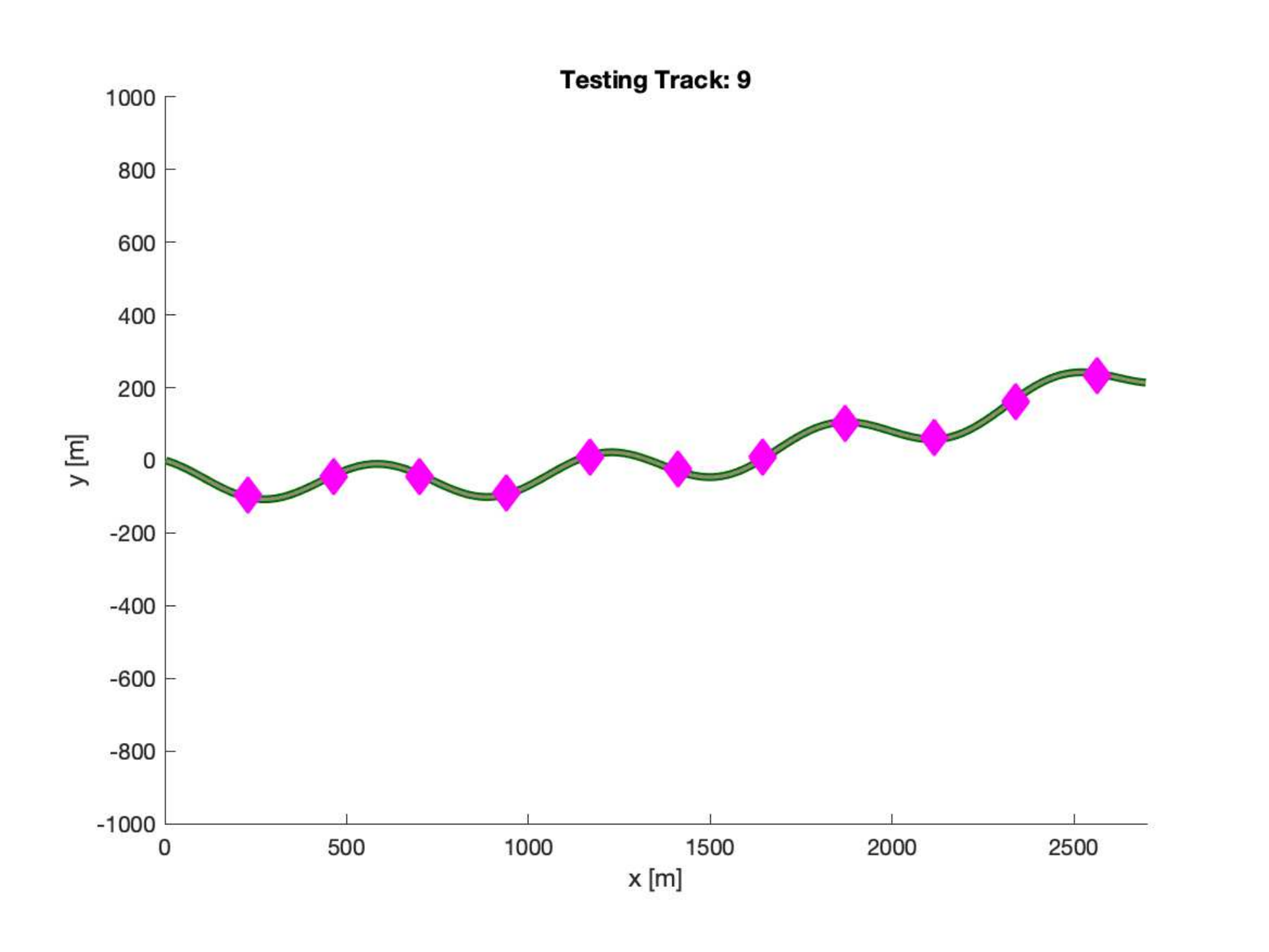}}
\subfloat[Track 10]{\includegraphics[width = 0.2\linewidth]	{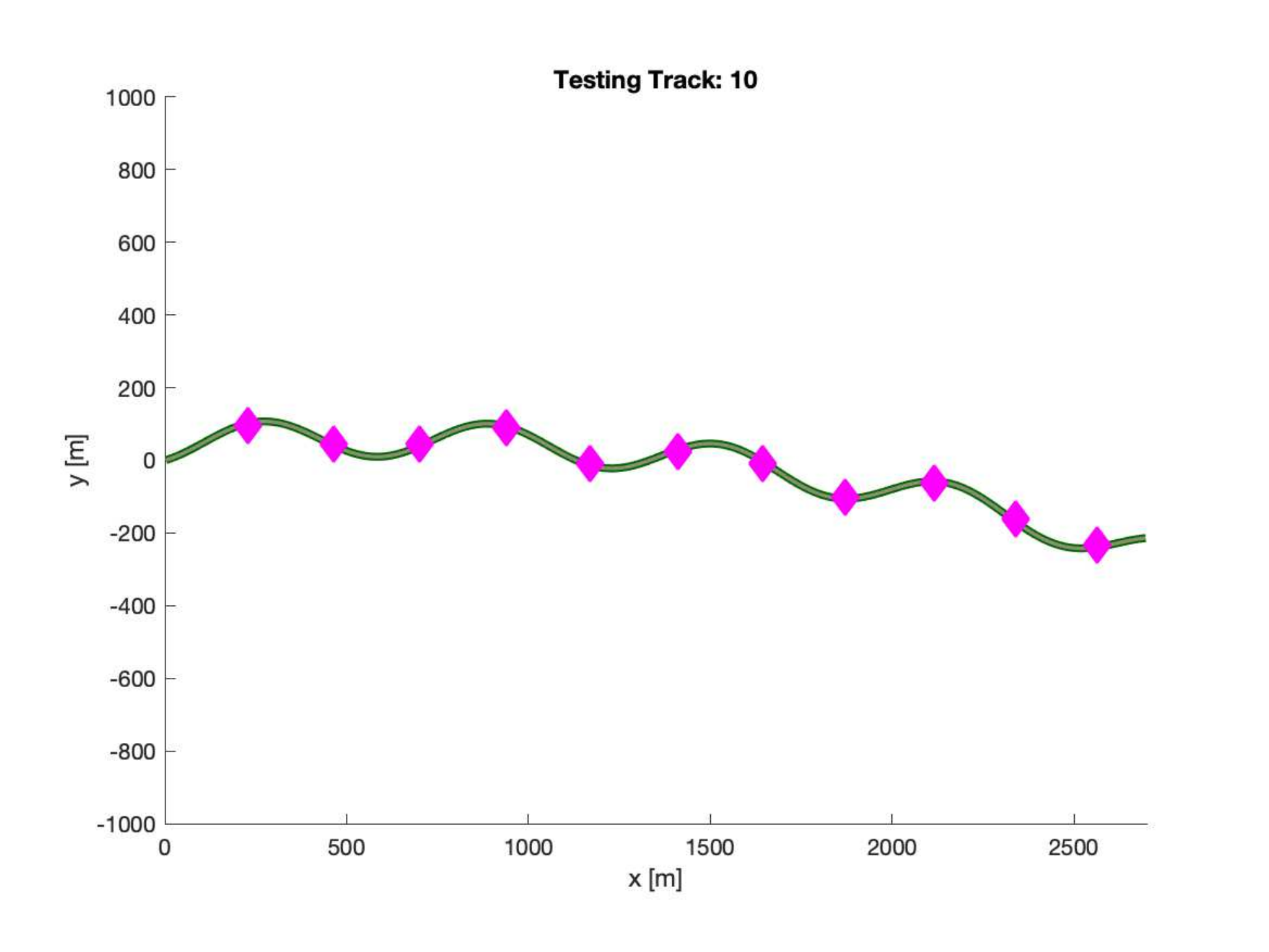}}
    \vspace{-2mm}
\caption{Candidate tracks. Magenta dots indicate the locations where the participants reported the difficulty of driving.}
\label{fig:candidate_tracks}
\end{figure}

The pilot study used a within-subjects design with 10 different candidate tracks (Fig.~\ref{fig:candidate_tracks}). The presentation of tracks followed a $10\times10$ Latin square design to eliminate potential order effects. %Figure~\ref{fig:candidate_tracks} shows the 10 candidate tracks.

% \textit{\textbf{Independent variables.}} The independent variable in the pilot study was ten different tracks. 
% The ten candidate tracks were designed to have small difficulty and the difficulty along the track is nearly constant. 

\subsubsection{Measures}
Along each track, participants reported the difficulty of driving at 11 locations using a 7-point Likert scale (1: easiest; 7: most difficult). After completing each track, participants also evaluated to what extent the track had the same difficulty anywhere along it (i.e. uniformity score), using another 7-point Likert scale (1: the same; 7: significantly different). 
% if the difficulty of driving is the same or significantly different at any location of the track using a 7-point Likert scale, where 1 represents "the same", 7 represents "significantly different". We named this measure as "uniform score" to represent the degree of uniformly distribution of the difficulty along the track.
For each track, the average of the 11 reported difficulty scores was calculated to represent the overall difficulty of the track (i.e. overall difficulty score). 

% and from the participant as the difficulty measure for each track. The magenta dots in Figure~\ref{fig:candidate_tracks} indicate the locations of the sign to notice the participants to report difficulties along the track.

% \subsubsection{Experimental procedure}
% Participants provided a signed informed consent and filled in a demographic survey. During the training session, the participants performed two trials on the training tracks, each took approximately 1.5 minutes. In the first trial, the participants only drove on the track and did not report difficulty. However, in the second trial, the participants drove on the track and reported difficulties at the 4 designed locations, indicated by a sign on the side of the road in the driving simulator. 

% In the official pilot study, the participants drove on 10 different tracks and reported difficulties at the 11 designed locations. After each track, the participants were asked to evaluate if the driving is the same or significantly different at any location of the track using 7-point Likert scale.

% After finishing all the ten trials, the subjects were required to fill a debriefing survey about outstanding questions and their opinions of or suggestions for the experiment they had just completed. 

\subsubsection{Result}
One-way repeated measures Analysis of Variance (ANOVA) was conducted with the driving track as the within-subjects variable. The results showed a non-significant difference between the ten tracks in their overall difficulty scores ($F(9, 81) = 1.161$, $p = 0.331$) and in their uniformity score ($F(9, 81) = 0.557$, $p=0.828$). Based on the results, we selected track 2, 3, 5, 6, 8, 9 to be used in Pilot Study 2 and Experiment 1, and track 2, 3, 6, 9 to be used in Experiment~2.
% \newpage

\vspace{-3mm}
\section{Pilot Study 2 -- Design of Surveillance Task}

We aimed to manipulate the difficulty of the dual-task mission and hence the human operators' workload by varying the surveillance task urgency. In Pilot Study 2, we selected the paces of the surveillance task, so that the difficulty and workload of the dual-task mission can be manipulated.

\subsubsection{Participants}
A total of 7 students participated in Pilot Study 2. Data of one participant was discarded due to equipment malfunction. The remaining 6 participants were on average 25.3 years old (\textit{SD} = 1.6 years) and had an average of 2.7 years of driving experience (\textit{SD} = 1.6 years). All participants had normal or corrected-to-normal vision.

\subsubsection{Apparatus and stimuli}
Pilot study 2 used the same dual-task shared control simulation platform as in Experiment 1. %Participants performed both the driving and surveillance tasks. 
The non-adaptive haptic shared control scheme was applied.

\subsubsection{Experimental design}
The pilot study used a within-subject design with six different time intervals of the surveillance task: $1.5$, $2.5$, $3.5$, $4.5$, $5.5$, and $6.5$~s, i.e. participants had to complete the detection task within any given time interval. The six time intervals were selected based on the results from our previous study \cite{luo2019toward}.  %Figure~\ref{fig:hist_response_time} shows the histogram of the response time of the participants doing the surveillance task.
Participants performed both the driving task and the surveillance task on 6 different tracks, each with a constant different time interval. The presentation of surveillance task conditions followed a $6 \times 6$ Latin square design to eliminate potential order effects. 

% time interval between stimuli
% surveillance task fixed time. 

% \textit{\textbf{Independent variables.}} The independent variable in the pilot study was six different surveillance task fixed time: $1.5, 2.5, 3.5, 4.5, 5.5, 6.5$ seconds.  

%\begin{figure}[h]
%    \centering
%    \includegraphics[width = 0.6\linewidth]{Figures/hist.eps}
%        \vspace{-3mm}
%    \caption{Histogram of response time for surveillance task from previous study. Red dash lines indicate 1.5, 2.5, 6.5~s respectively.} 
%    \label{fig:hist_response_time}
%\end{figure}

\subsubsection{Measures}
Participants reported their workload of the dual-task mission using the NASA TLX survey \cite{hart1988development} and their perceived difficulty of the dual-task mission. 

% \subsubsection{Experimental procedure}
% Participants provided a signed informed consent and filled in a demographic survey. After that, they were provided instructions and training. The training session is the same as Experiment 1 except the surveillance task fixed time is $5.5, 3.5, 1.5$ seconds instead of $6.5, 2.5, 1.5$ seconds.

% During the official experiment, participants performed driving task and surveillance task on 6 different tracks with 6 different surveillance task fixed time. Each track took approximately 3 minutes. Unlike Experiment 1 and 2, the surveillance task fixed time was constant during each track. After each trial, the participants were asked to fill a post survey about the workload and difficulty during each track.

% After finishing all the six trials, the subjects were required to fill a debriefing survey about outstanding questions and their opinions of or suggestions for the experiment they had just completed. 

\subsubsection{Result}
One-way repeated measures Analysis of Variance (ANOVAs) was conducted with the surveillance time interval as the within-subjects variable. The results showed a significant difference of time interval on workload ($F(5,25) = 10.458$, $p<0.001$) and difficulty ($F(5,25) = 13.423$, $p<0.001$). We then performed a series of $t$ tests between different pairs of time intervals. %The results revealed significant differences in workload and difficulty between $1.5$~s and other time intervals (Table~\ref{tab:pilot_2_result}).
The results revealed significant differences in workload and difficulty between $1.5$ and $2.5$~s (workload: $p<.001$, difficulty:$p=.006$), between $1.5$ and $3.5$~s (workload: $p=.005$, difficulty:$p=.012$), between $1.5$ and $4.5$~s (workload: $p =.004$, difficulty:$p=.006$), between $1.5$ and $5.5$~s (workload: $p=.001$, difficulty:$p<.001$), and between $1.5$ and $6.5$~s (workload: $p=.004$, difficulty:$p<.001$). 
The differences between any other pairs of time intervals were non-significant.

%\begin{table}[]
% \centering
% \caption{Series of $t$ tests between different pairs of time interval}
%     \vspace{-3mm}
% \label{tab:pilot_2_result}
% \begin{adjustbox}{width=0.9\linewidth}
% \begin{tabular}{cccccc}
% \hline
%           & 1.5s vs. 2.5s & 1.5s vs. 3.5s & 1.5s vs 4.5s & 1.5s vs. 5.5s & 1.5s vs. 6.5s \\ \hline
% Workload   & $p < .001$    & $p = .005$    & $p = .004$   & $p = .001$    & $p = .004$    \\ \hline
% Difficulty & $p = .006$    & $p = .012$    & $p = .006$   & $p < .001$    & $p < .001$    \\ \hline
% \end{tabular}
% \end{adjustbox}
% \end{table}

Based on the results, we selected $1.5$- and $6.5$-second time intervals to be used in the Experiment 1 and Experiment 2 to induce varying levels of workload. Note in Experiment 1, we also included the $2.5$-second time interval, as we were interested to explore participants' performance with a slightly larger time interval compared to the $1.5$-second time interval.
% for surveillance task fixed time to control the participant's workload on the dual task. {\color{red}{However, we also involved $2.5$ second in Experiment 1 to collect data for future study of workload estimation in different levels.}}

% \newpage
%    \vspace{-3mm}
%\section{Scales Used to Measure Workload and Trust}
%\vspace{-10mm}
%\begin{figure}[h]
%    \centering
%    \includegraphics[width = 0.9\linewidth]{Figures/workload_trust.PNG}
%    \label{fig:Scale_measure_workload_trust}
%\end{figure}

\bibliographystyle{IEEEtran}

\bibliography{mybib}

    \vspace{-15mm}
\begin{IEEEbiography}[\vspace{-5mm}{\includegraphics[width=1in,clip,keepaspectratio]{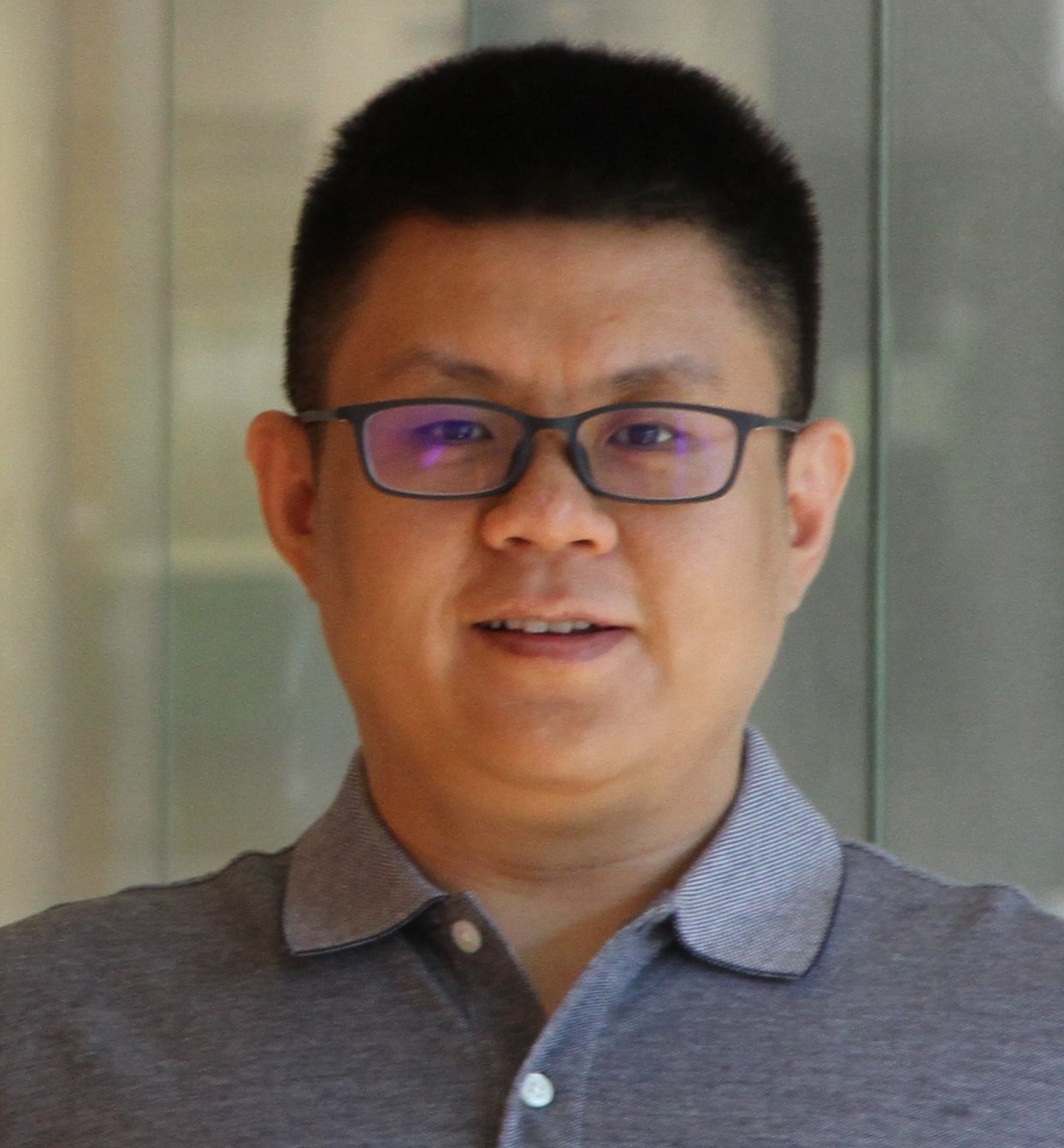}}]{Ruikun Luo}
is a Ph.D. candidate at the Robotics Institute, University of Michigan, Ann Arbor. Prior to joining the University of Michigan, he obtained a M.S. in Mechanical Engineering from Carnegie Mellon University in 2014 and a B.S. in Mechanical Engineering and Automation from Tsinghua University, China in 2012. His research interest is human-robot interaction.
\end{IEEEbiography}

    \vspace{-17mm}
\begin{IEEEbiography}[\vspace{-5mm}{\includegraphics[width=1in,clip,keepaspectratio]{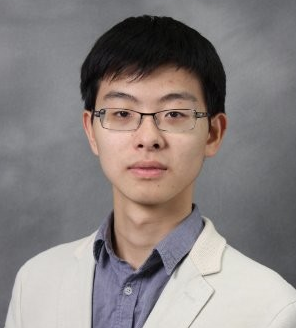}}]{Yifan Weng}
is a Ph.D. candidate at the Mechanical Engineering, University of Michigan, Ann Arbor. He received the B.S.E. degree in mechanical engineering from Shanghai Jiao Tong University, Shanghai, China and from Purdue University, IN, US in 2016 and the M.S.E. degrees in mechanical engineering from the University of Michigan, Ann Arbor, MI, USA, in 2018.
\end{IEEEbiography}

    \vspace{-17mm}
\begin{IEEEbiography}[\vspace{-8mm}{\includegraphics[width=1in,clip,keepaspectratio]{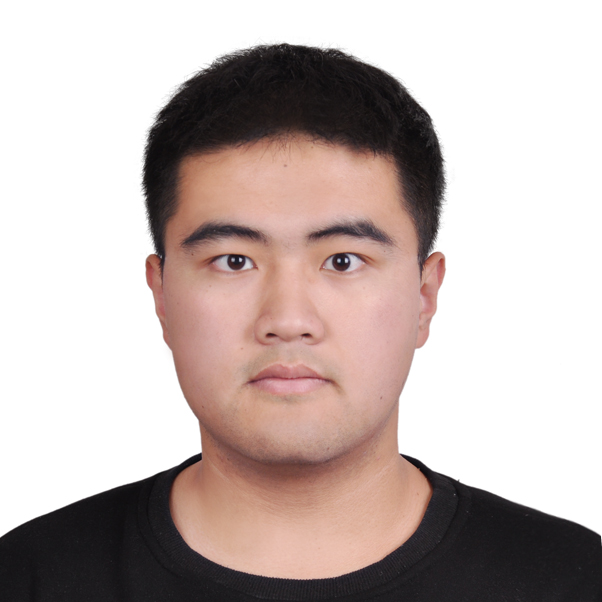}}]{Yifan Wang}
is a Ph.D. student at the electrical engineering and computer science, University of Michigan, Ann Arbor. He received the B.S. degree in electrical engineering from Xi'an Jiao Tong University, China, in 2018, and the M.S.degree in electrical and computer engineering from the University of Michigan, Ann Arbor, in 2019.  His research interests include human-computer interaction, medical imaging processing and reinforcement learning theory.
\end{IEEEbiography}

    \vspace{-17mm}
\begin{IEEEbiography}[{\includegraphics[width=1in,height=1.25in,clip,keepaspectratio]{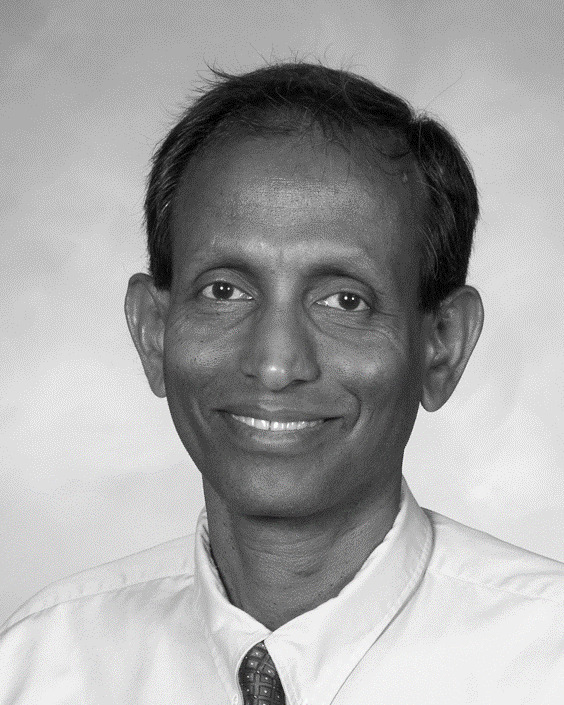}}]{Paramsothy Jayakumar}
received his M.S. and Ph.D. degrees in structural dynamics from Caltech, and B.Sc. Eng. from the University of Peradeniya, Sri Lanka. He is a Senior Research Scientist, SAE Fellow, and a member of the Analytics Team at the U.S. Army Ground Vehicle Systems Center (GVSC) in Warren, Michigan. He is a member of the U.S. Army Acquisition Corps, an Honorary Fellow of the Department of Mechanical Engineering at the University of Wisconsin Madison, and an Associate Editor for the ASME Journal of Computational and Nonlinear Dynamics.
\end{IEEEbiography}

    %\vspace{-15mm}
\begin{IEEEbiography}[{\includegraphics[width=1in,height=1.25in,clip,keepaspectratio]{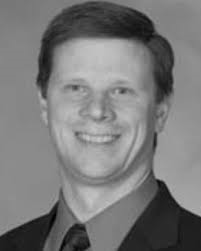}}]{Mark Brudnak}
(M’04) received the B.S. degree in electrical engineering from Lawrence Technological University, Southfield, MI, in 1991, and the M.S.degree in electrical and computer engineering and the Ph.D. degree in systems engineering from Oakland University, Rochester, MI, in 1996 and 2005, respectively. He is currently an Associate Director at the U.S. Army Ground Vehicle Systems Center (GVSC), Warren, MI. In this capacity, he oversees the operation of laboratories for durability testing, vehicle characterization, and human-in-the-loop motion base simulation.
\end{IEEEbiography}
    \vspace{-17mm}

\begin{IEEEbiography}[{\includegraphics[width=1in,height=1.25in,clip,keepaspectratio]{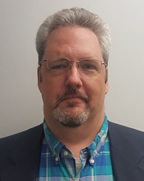}}]{Victor Paul}
 serves as a Team Leader in the Ground Vehicle System Center’s System Ground Vehicle Simulation Laboratory (GVSL) where he has worked for 28 years. He holds extensive knowledge in the area of motion base simulation and its application in both man and hardware in the loop experiments. He is currently a senior advisor for the Crew Optimization and Augmentation Technologies Science and Technology program and is supporting the development of Crew Station and Squad interfaces for the Next Generation Combat Vehicle Mission Enabling Technology Demonstrator.
\end{IEEEbiography}
    \vspace{-17mm}

\begin{IEEEbiography}[{\includegraphics[width=1in,height=1.25in,clip,keepaspectratio]{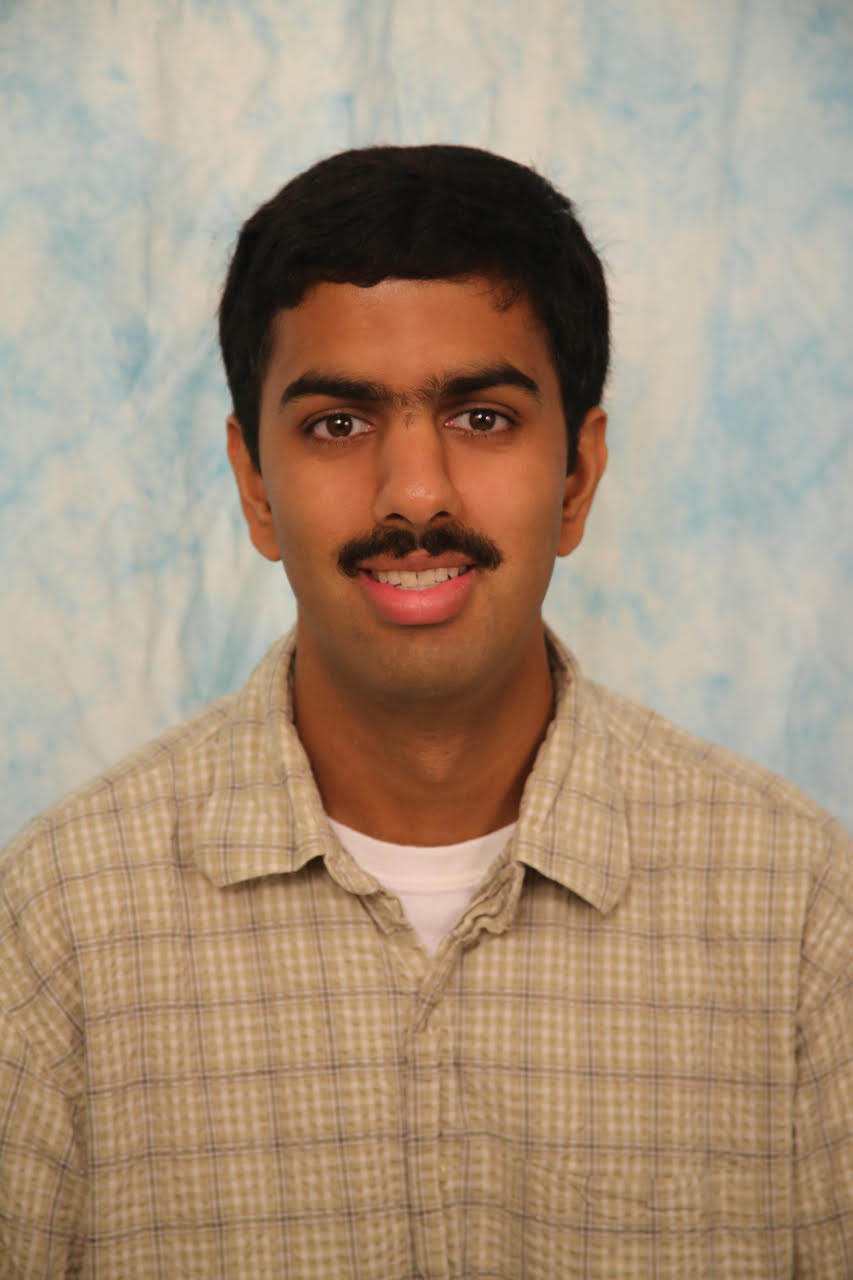}}]{Vishnu R. Desaraju}
 received the B.S.E. degree in Electrical Engineering from the University of Michigan, Ann Arbor, MI in 2008, S.M. degree in Aeronautics and Astronautics from the Massachusetts Institute of Technology, Cambridge, MA in 2010, and M.S. and Ph.D. degrees in Robotics from Carnegie Mellon University, Pittsburgh, PA in 2015 and 2017, respectively. He is currently a Research Scientist at the Toyota Research Institute, Ann Arbor, MI developing automated driving technologies. His research interests include real-time motion planning and control of constrained, uncertain systems.
\end{IEEEbiography}
    \vspace{-17mm}

\begin{IEEEbiography}[{\includegraphics[width=1in,height=1.25in,clip,keepaspectratio]{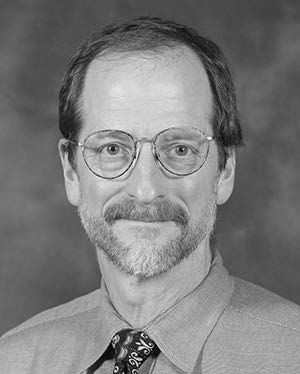}}]{Jeffrey L. Stein}
received the B.S. degree in premedical studies from the University of Massachusetts, Amherst, MA, in 1973, and the S.B., S.M, and Ph.D. degrees in mechanical engineering from the Massachusetts Institute of Technology, Cambridge, MA, in 1976, 1976, and 1983, respectively. Since 1983 he has been with the University of Michigan, Ann Arbor, MI, where he is currently a Professor of Mechanical Engineering. His research interests include computer based modeling and simulation tools for system design and control, with applications to vehicle-to-grid integration, vehicle electrification, conventional vehicles, machine tools, and lower leg prosthetics.
\end{IEEEbiography}
    \vspace{-17mm}

\begin{IEEEbiography}[{\includegraphics[width=1in,height=1.25in,clip,keepaspectratio]{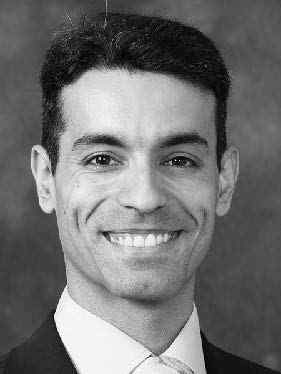}}]{Tulga Ersal}
received the B.S.E. degree from the Istanbul Technical University, Istanbul, Turkey, in 2001, and the M.S. and Ph.D. degrees from the University of Michigan, Ann Arbor, MI USA, in 2003 and 2007, respectively, all in mechanical engineering. He is currently an Associate Research Scientist in the Department of Mechanical Engineering, University of Michigan, Ann Arbor, MI. His research interests include modeling, simulation, and control of dynamic systems, with applications to vehicle and energy systems.
\end{IEEEbiography}
    \vspace{-17mm}

\begin{IEEEbiography}[{\includegraphics[width=1in,height=1.25in,clip,keepaspectratio]{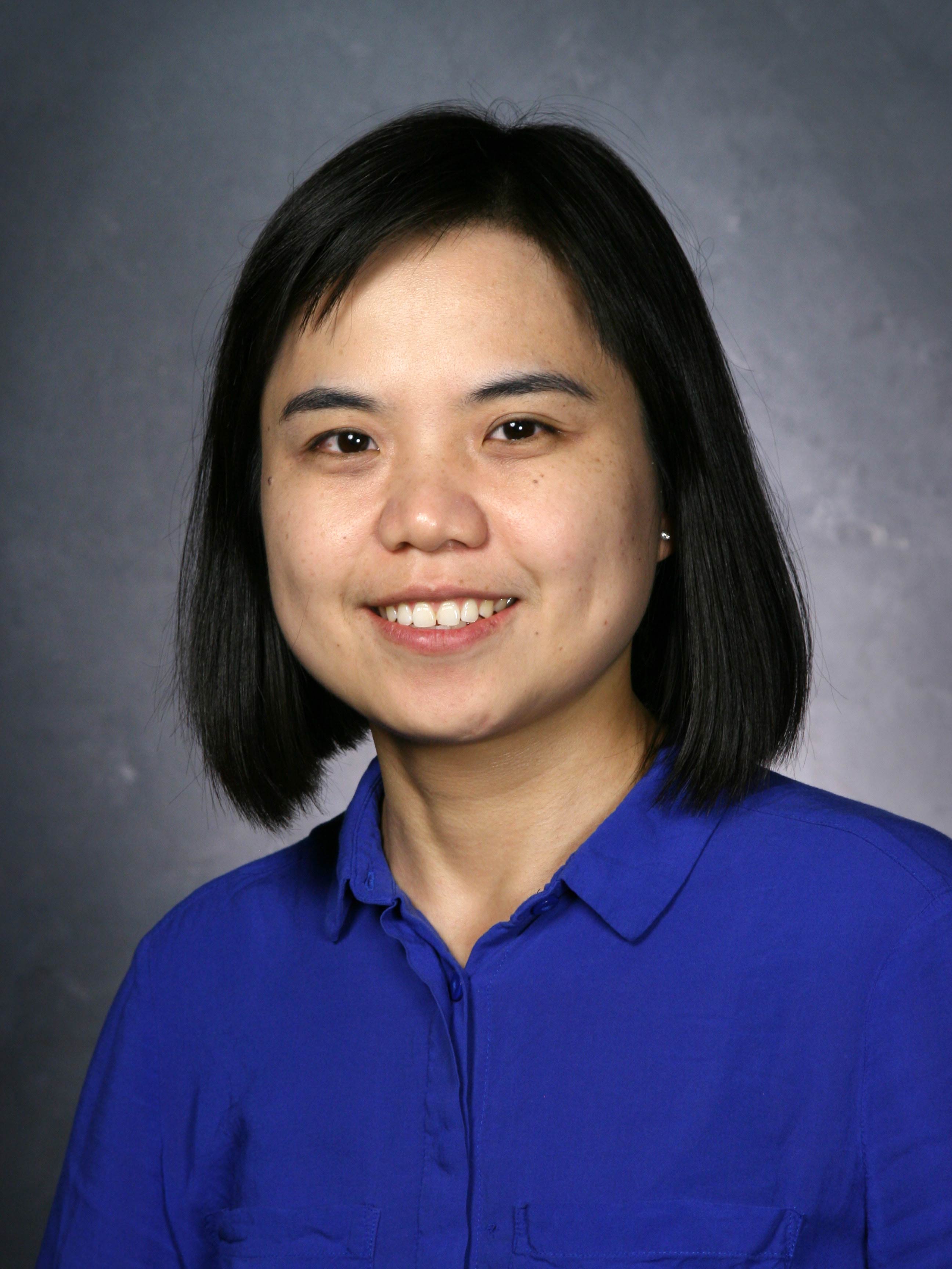}}]{X. Jessie Yang}
is an Assistant Professor at the Department of Industrial and Operations Engineering, University of Michigan Ann Arbor. She obtained her PhD in Mechanical and Aerospace Engineering (Human Factors) from Nanyang Technological University, Singapore in 2014.
\end{IEEEbiography}

% \newpage
% \section{Biographies}
% \textbf{Author 1} is a\\

% \textbf{Author 2} is a\\

% \textbf{Author 3} is a\\

% \textbf{X. Jessie Yang} is an Assistant Professor in the Department of Industrial and Operations Engineering at the University of Michigan Ann Arbor. She obtained a PhD in Mechanical and Aerospace Engineering (Human Factors) from Nanyang Technological University Singapore in 2014.\\

\end{document}